# B-site spin-state and anti-site disorder driven multiple-magnetic phases: Griffiths phase, re-entrant cluster glass and exchange bias in double perovskite $Pr_2CoFeO_6$

Arkadeb Pal[1], P. Singh[1], Shiv Kumar[3], S. Ghosh[1], Sudip Kumar Saha[4], Manoranjan Kumar[4], Amitabh Das[5], Eike F. Schwier[3], Masahiro Sawada[3], Kenya Shimada[3], A. K. Ghosh[2], Sandip Chatterjee[1#]

[1]Indian Institute of Technology (BHU) Varanasi 221005, India

[2]Banaras Hindu University, Varanasi 221005, India

[3]Hiroshima Synchrotron Radiation Center, Hiroshima University, Kagamiyama 2-313, Higashi-Hiroshima 739-0046, Japan

[4]S. N. Bose National Centre for Basic Sciences, Kolkata 700098, India

[5]Bhaba Atomic Research Centre, Mumbai 400085, India

Corresponding Email: schatterji.app@iitbhu.ac.in

## Abstract

We report the comprehensive experimental results identifying the magnetic spin ordering and the magnetization dynamics of a double perovskite $Pr_2CoFeO_6$ by employing the (dc and ac) magnetization, powder neutron diffraction (NPD) and X-ray magnetic circular dichroism (XMCD) techniques. X-ray diffraction and neutron diffraction studies revealed that $Pr_2CoFeO_6$ adopts a B-site disordered orthorhombic structure with space group Pnma. Additionally, ab initio band structure calculations performed on this system suggested an insulating anti-ferromagnetic (Fe-Fe) ground state. Magnetometry study showed the system to possess a spectrum of interesting magnetic phases including long range antiferromagnetic (canted) spin ordering ($T_N$ ~269 K), Griffiths phase, re-entrant cluster glass (RCG) ($T_G$~ 34 K) and exchange bias. However, the NPD study divulged the exhibition of a long range G-type (below $T_N$ ~269 K) of spin ordering by Fe spins. Spin dynamics study by ac susceptibility technique confirmed the system possessing long range ordering at higher temperatureundergoes a RCG transition at ~34 K. Existence of Griffiths phase was confirmed by non-analytic field variation of magnetization and Heisenberg type temporal spin relaxation above long range ordering temperature $T_N$ ~269 K. The anti-site disorder related to the B-sites (Co/Fe) is found to be the main driving force forthe observed multiple magnetic phases. Furthermore, the electronic structure probed by the X-ray absorption spectroscopy (XAS) study suggested a nominal valance state of +3 for both of the B-site ions (Co/Fe) which in turn triggered the anti-site disorder in the system. Magnetic, XRD, NPD and XAS analysis yielded a low spin state (LS) for the $Co^{3+}$ ions. The random non-magnetic dilution of magnetic $Fe^{3+}$ (HS) ions by $Co^{3+}$ (LS) ions essentially played a crucial role in manifesting the magnetic properties of the system.

## Introduction

Materials that give responses to various external stimuli gained much interest due to their intriguing rich physics and prospect for technological device applications [1-4]. Particularly, the class of oxide double perovskites $A_2BB^{'}O_6$ (A= Rare earth ions or alkaline ions; B/B$^{'}$= transition metal ions) with rock salt ordered structure [5] has attracted a great deal of research attention due to their diverse exotic properties including giant magneto-resistance[6-7], spin reorientation [8], cationic ordering [9], magnetocaloric effects [10], colossal magneto-dilectric effect [11], E-type (↑↑↓↓) of ordering driven ferroelectricity [12], metamagnetic transition[13-14], anti-site disorder driven multi-glass phases [15], giant exchange bias [16], Griffiths phase [17-18] etc. Hence, these complex and interesting physical properties can be harnessed to fabricate innovative devices for practical applications. The structure of double perovskite consists of double typical $ABO_3$ perovskite unit cells, where the two different B and B$^{'}$ atoms are forming rock-salt type ordering (checker board pattern). Most of the Ni/Co/Fe (B) – Mn (B$^{'}$) based ordered double perovskites are ferromagnetic insulators possessing high temperature magnetic transition owing to the $180^0$ FM super-exchange interactions between $B^{2+}$ and $B^{'4+}$ ions (half filled d orbital) which is best understood by Goodenough-Kanamori rules[11,19-20]. For double perovskites, anti-site disorder i.e. an interchange of B/B$^{'}$ sites is well known to have profound effects on its physical properties, particularly on its magnetic properties which calls for rigorous theoretical and experimental investigations[14-15,19-22]. Eventually, anti-site disorder can cause sizeable deviations from ferromagnetism by introducing additional antiferromagnetic clustered regions via superexchange interactions in the form of $B^{2+}$-$O^{2-}$-$B^{2+}$ and $Mn^{4+}$-$O^{2-}$-$Mn^{4+}$; which in turn result in introduction of competition between FM and AFM interactions [15]. It is well established that competing FM and AFM interactions are the basic ingredients in anticipating emergence of short range ordering related secondary magnetic phases like low temperature spin-glass, exchange bias and Griffiths phase etc [15-18]. In the widely studied systems $La_2NiMnO_6$ and $La_2CoMnO_6$, the role of anti-site disorder in the evolution of multiple magnetic phases separated by antiphase domains have been extensively studied [15,19]. Antisite disorder was seen to play crucial role in emergence of spin-glass behaviour as well as enhancing magneto-electric coupling in the system $La_2NiMnO_6$[15]. In another extensively studied Y-based double perovskite compound $Y_2CoMnO_6$, the antisite disorder has been seen to play major role in deciding its magnetic properties [13-14]. So far, it was believed that $Y_2CoMnO_6$ shows ferroelectricity owing to its E-type (↑↑↓↓) of Co/Mn magnetic ordering [23]. However, J. Blasco et.al have experimentally shown in details how the different degree of antisite disorder affects its magnetic as well as electrical properties [14]. In contrast to the widely investigated $R_2BMnO_6$ (R=La,Y,Lu, Pr, Sm, Dy,Tb,Ho etc and B=Co, Ni etc) compounds, the studies on the Fe based double perovskites i.e. $R_2BFeO_6$ oxides, are comparatively limited and thus there are much more opportunities to explore their diverse interesting physical properties. It is reported that the $B^{3+}$ and $B'^{3+}$ ions are usually raise antisite disorder by random site distribution in the octahedral sites, thus leading to orthorhombic (Pnma) or rhombohedral symmetry [8,24-25]. Hence, in systems $R_2BFeO_6$, the $B^{3+}$ and $Fe^{3+}$ ions cause appreciable antisite disorder which has strong potential leading to various extraordinary properties as discussed above. Additionally, different compounds with A site occupying a magnetically active rare earth (4*f*) ion $R^{3+}$, show wide spectrum of interesting phenomena due to the additional competing 4*f*-3*d* negative exchange interactions owing to the localized and much more complex configuration of the 4*f* orbitals relative to the transition metal 3*d* orbitals [26-29]. For example, in various orthoferrites $RFeO_3$ (R=Er, Sm, Ho, Dy, Tb, Nd, Pr, etc) and very recently in a double perovskite compound $Ho_2CoFeO_6$, spin-reorientation transitions have been reported and the underlying physics wasunderstood bythe competition between Zeeman energy and the magnetic anisotropy [8, 30-34]. Here the magnetic anisotropy is lead by the competing complex interactions 3*d*-3*d* , 4*f*-3*d* and 4*f*-4*f* consisting of isotropic, anisotropic symmetric and anti-symmetric super-exchange interactions. In a similar orthoferrite $Dy_{0.5}Pr_{0.5}FeO_3$, field induced two fold spin reorientation (SR) transition ($\Gamma_4 \rightarrow \Gamma_1 \rightarrow \Gamma_4$) was reported recently where the intriguing physics involved was ascribed to the effective anisotropic field in the system raised by the mutual interactions between Dy-4*f* and Pr-4*f* electrons and their competing interactions (4*f*-3*d*) with the $Fe^{3+}$ (3*d*) sublattices [34]. Another interesting phenomenon observed in orthoferrites $RFeO_3$, is the evolution of weak ferromagnetism raised from canted $Fe^{3+}$ spins due to the spin–orbit coupling induced antisymmetric exchange interactions which is described by Dzyaloshinskii, Moriya, and Treves in the dominant antiferromagnetic background [35-37]. On the other hand, the rare earth based cobaltite oxides $RCoO_3$, is well-known systems since 1950s and particular attention has been given to the thermally driven spin state transition from the low spin LS ($t_{2g}^6$) state to the higher spin states of the $Co^{3+}$ ions[38-43]. However, it is still remained debated whether the spin state transition occurs directly to a high spin state (HS $t_{2g}^4 e_g^2$) or to an intermediate state (IS $t_{2g}^5 e_g^1$)and lot of research works have been devoted to this [40-41]. In particular for $PrCoO_3$, it is controversial whether the spin state of $Co^{3+}$ is in LS or higher states (IS or HS) upto 300K [42]. In contrast to paramagnetic bulk $PrCoO_3$, in its epitaxial thin film, $Co^{3+}$ (HS) long range ferrimagnetic ordering has been reported [43]. Thus, the spin state transition in $PrCoO_3$ got renewed interest so as to get an insight into the underlying correlated electron properties and competing degrees of freedom determining the spin state. Hence, realizing the potential to give rise to many interesting physical properties as discussed above, the replacement of Mn by Fe in double perovskite family can be of particular scientific interest. Therefore, with the aim of giving a comprehensive study of Co/Fe interactions driven magnetic ground state and the role of ASD in deciding the physical properties in Co/Fe based systems, we synthesized the double perovskite system $Pr_2CoFeO_6$ (PCFO) and carried out detailed investigations on its magnetic, structural, electrical and electronic properties and presented it in this report. Here, the comparable ionic radii and same nominal charge states (both +3 for Co and Fe) and the strong interactions between magnetic $Pr^{3+}$ (4*f*) with Co/Mn (3d) sublattices along with the antisite disorder are expected to trigger exotic magnetic phenomena.

In this paper, we have presented results from suit of experimental measurements comprising temperature dependent (DC and AC) magnetization measurements, X-ray diffraction, X-ray absorption spectroscopy (XAS), X-ray magnetic circular dichroism (XMCD) and Neutron diffraction study of PCFO.

## I. EXPERIMENTAL DETAILS

### (a) Material Synthesis:

The polycrystalline $Pr_2CoFeO_6$ sample used in the present investigation was prepared following the standard conventional solid state reaction method. The high purity (>99.99%) oxide powders $Pr_6O_{11}$, CoO and $Fe_2O_3$ as precursors were weighted in proper molar ratios and then intimately ground for 1hour in a mortar. The thoroughly ground mixture was subjected to an initial heat treatment at $1000^0$ C for 24 hours in air. The resulting powder was then reground and was again subjected to several heating cycles at $1200^0$C with intermittent grinding and reheating steps for several days. In the final step, the resulting powder thus obtained was pressed into pellets and sintered at $1300^0$C for 36 hours followed by a slow cooling ($0.5^0$C/min) to room temperature.

### (b) Material characterization:

The phase purity of the samples was checked by powder X-ray diffractogram (XRD) obtained by a Rigaku Miniflex II X-ray diffractometer (Cu $K_\alpha$) and was refined by Rietveld method using FULLPROF suite software. The Neutron diffraction studies were carried out by a neutron powder diffractometer ($\lambda$ =1.2443A°) having five position sensitive linear detectors at Dhruva reactor stationed at Bhaba Atomic Research Centre, Trombay, India. The superconducting quantum interference device (SQUID) based magnetic property measurement system (Quantum Design-MPMS) was employed for all the temperature dependent magnetization measurements. The XAS and XMCD measurements were performed at the BL14 beamline of Hiroshima Synchrotron Radiation Centre, Hiroshima University, Japan. In recording the spectra, total electron yield (TEY) mode has been used as it requires relatively easy experimental setup and gives high signal to noise ratio. A base pressure of $4\times10^{-8}$ Pa was maintained in the experimental chamber where the sample was mounted. The photon energy range of the beamline was 400-1200 eV which is compatible for XAS study at $L_{2,3}$ edges of Co and Fe (3d transition metals).

### (c) Computational details :

We have performed our study based on density functional theory (DFT) using Vienna ab initio simulation package (VASP). Exchange-correlation potential (Perdew-Burke-Ernzerhof exchange-correlaton functional) is approximated with generalized gradient approximation (GGA). The projector augmented wave method (PAW) is used for core-valence interaction. The calculations are performed with K-mesh of 8×5×8 for $Pr_2CoFeO_6$ with Pnma space group. We have considered plane-wave basis up to cut-off energy 600 eV for convergence. The lattice parameters are optimized before the calculation of DOS to reduce internal forces. To see the spin polarized partial and total DOS, we have considered the on-site coulomb correction (GGA+U).

## II. RESULTS AND DISCUSSIONS

### A. X-ray Diffraction Study:

The crystallographic information has been extracted by refining the XRD data using FULLPROF program suite. The XRD pattern recorded at 300 K along with its Rietveld refinement is shown in Fig. 1. The inset is showing the pictorial diagram of the crystal structure. The refinement suggests that the compound crystallizes in disordered orthorhombic phase with symmetry Pnma, thus it in turn indicates random distribution of Co/Fe ions at B-sites. All the peaks are indexed according to the orthorhombic structure (a×c×b : $(2)^{1/2}a_c\times 2a_c\times(2)^{1/2}a_c$, here $a_c$= 3.843Å, being the lattice constant of perovskite sub-cell) within Pnma symmetry. No trace of chemically impure phase is found, suggesting our sample to be of single phase. Factually, the random (B-site) cationic distribution in PCFO can be understood by the means of same charge states of Co/Fe ions (+3), as the ordered rock-salt type arrangement of B-cations leading to a monoclinic structure $P2_1/n$, requires the charge difference between B and $B^{'}$ [28]. Andersen et. al. have investigated the effects of having same charge state of B-site ions and shown how it affects its structure[44]. One of its important aspects is that the structure becomes centro-symmetric due to the random site distribution of Co/Fe ions. The refinement suggests that PCFO sample also crystallizes in a centro-symmetric orthorhombic structure with Glazer notation $a^+a^-b^-$ tilt system [45]. For, PCFO system, the deviation from cubic to orthorhombic structure is triggered by the small size of A-site ion. However, the distortion of octahedra: Co(or Fe)$O_6$ can simply be measured by the formula $\delta=(180^0-\varphi)/2$, where the φ is a measure of angle Co(Fe)-O-Co(Fe) [46]. Here, the value of δ is $7.715^0$, which clearly suggests the presence of sizeable distortion in the octahedral. Again, for a random cationic distribution in the B-sites, the average bond-lengths of Co and Fe with O (Co-O and Fe-O) should be almost uniform while that for a perfectly ordered system show appreciable differences [47]. Eventually, the detailed structural investigations on the bond-lengths and bond-angles, reveal the average bond-lengths Co/Fe-O to bequite similar (Table. 1), thus clearly indicating towards the presence of random distribution of Co/Fe at B-sites. Again, the ionic radii for $Fe^{3+}$ (H.S), $Co^{3+}$ (LS) and $O^{2-}$ are 0.645Å , 0.545Å, 1.38Å respectively, hence simply by summing up their ionic radii and taking the mean, we get the average theoretical bond-length Fe/Co-O to be 1.97Å [47]. Thus, it shows close match with the bond-length Fe/Co-O(2)=1.96 Å which was extracted from refinement of XRD data (Table-1). Now, for $Co^{3+}$(HS) ( 0.61 Å) and $Fe^{3+}$ (H.S), the average bond length can similarly be found : Fe/Co-O=2.01 Å which does not fit with none of the experimentally obtained bond length Fe/Co-O(1) or Fe/Co-O(2). Hence, XRD analysis suggests $Co^{3+}$ to be in low spin state (LS) in PCFO.

## B. Electronic and magnetic properties study by ab initio calculations:

We have performed the ab initio calculations based on density function theory (DFT) for PCFO to get more insights into its electronic and magnetic structures. The structure has been optimized with orthorhombic Pnma symmetry (where, the ionic positions of the atoms were optimized keeping the shape and volume of the unit cell fixed). The structure was relaxed till the Feynman-Hellman forces were reduced below 0.001 eV $(A^0)^{-1.}$ The optimized structure reached to the lowest energy of ~ -154.421 eV.

After the structural optimization, we have used this Pnma structure with lowest energy to calculate the density of states (DOS). All the DOS calculations have been carried out with generalized gradient approximation (GGA) scheme for the exchange correlation potential (i.e. using GGA+U approximation). The calculations have been done with Hubbard U correction i.e. $U_{eff}$=U-J (here J and U are exchange and Coulomb parameters respectively) which are considered to be ~6 eV for Pr-4f states [48], ~6 eV for Co-3d states [49] and ~4 eV for Fe-3d states[50]. We have performed our calculations both for ferromagnetic and anti-ferromagnetic couplings among Fe spins (since $Co^{3+}$ ions are non-magnetic in its ground state). However, the calculations yielded that the structure with anti-ferromagnetic coupling among Fe spins has the less energy (~ -148.07 eV) as compared to that for ferromagnetic coupling (energy ~ -147.558 eV). Hence, the calculation predicts an anti-ferromagnetic ground state for the present PCFO system as the anti-ferromagnetic interactions are energetically favourable. Eventually, small energy difference between the structures with these two couplings suggests that ferro-magnetic contribution can also be there at finite temperatures. We have calculated the total density of states (TDOS) for PCFO system for the anti-ferromagnetic interactions between Fe spins. Fig. 2 (a) depicts the TDOS as a function of energy (which is scaled against the Fermi energy). The splitting between the up and down spin bands can be observed in the TDOS pattern which is due to the octahedral distortion present in the system. Moreover, from the Fig. 2 (a), the band gap is estimated for the up spin band to be ~1.5032 eV whereas that for the down spin band is found to be ~ 1.231 eV. Absence of the DOS at the Fermi level and presence of high band gap clearly suggest the system to be insulating in nature. Interestingly, the resistivity measurement of PCFO at room temperature showed a value of ~886 Ohm-m which confirmed the systems insulating behaviour. Thus the TDOS calculation corroborates with the experimental results.

To estimate different contributions from different states e.g. Pr –f/d/p/s, Co-d/p/s, Fe-d/p/s and O-p/s towards the TDOS, we have calculated partial density of states (PDOS). Fig. 2(b-e) are showing the up and down spin integrated PDOS for Pr-s/p/d/f, Co-s/p/d, Fe-s/p/d and O-s/p states. It is evident from Fig. 2(b-e) that Pr-f, Co-d, Fe-d and O-p states have the dominant contribution in their respective PDOS. No finite DOS is available near the Fermi level for none of the calculated PDOSs, thus confirming the insulating nature of the system. A large splitting can be observed in Pr-f PDOS spectra which leads to large energy gap between the unoccupied and occupied states. This large splitting also indicates that the Pr-f electrons are highly localized. Eventually, the large energy gap in Pr-f states strongly affects the Co/Fe-3d and O-2p states. These Co/Fe-3d and O-2p states get adapted by Pr-4f symmetry and hence PDOS related to these states appear in the same energy range as that of Pr-4f PDOS. Fig. 2(f) is showing the PDOS in both the spin channels for Pr-f, Co-d, Fe-d and O-p states. It is clear from the PDOS curves that there is significant hybridization among Co/Fe-3d and O-2p states. It is also evident that in down spin channel both the Fe-3d and Co-3d have mostly unoccupied states. On the other hand, in the up spin channel, most of the Fe-3d and Co-3d states are occupied. The Fig. 2(f) also suggests that in its ground state, Pr-f states will also contribute towards spin polarization. The asymmetric nature in the spin resolved PDOS of Fe-3d states clearly suggests its magnetic contribution in its ground state. However, small spin polarization observed for Co-3d and O-2p is due to strong hybridization with Fe-3d and Pr-4f states.

## C. Neutron Diffraction Study:

To get an insight into the microscopic spin arrangement as well as structural order in PCFO, we have undertaken neutron powder diffraction (NPD) study at two different temperatures 300 K and 6 K. The neutron thermo-diffractograms along with its Rietveld refinements are shown in Fig. 3(a-b). Eventually, we know that the B-site ordered double perovskites crystallize in monoclinic $P2_1/n$ space group which requires a minimum charge state difference of +2 in between two B-site ions. However, understanding the fact that Co and Fe have the same nominal charge states of +3, it is expected that there will be random B-site distribution of $Co^{3+}$ and $Fe^{3+}$ ions, thus giving rise to anti-site disorder. Again, the large difference in the coherent neutron scattering lengths of Co (2.49 fm) and Fe (9.45 fm) allows us to probe the degree of B-site structural ordering in the system. Hence, NPD study has been done to precisely know if there is anti-site disorder present in the system PCFO. We have attempted to fit the NPD data by monoclinic $P2_1/n$ symmetry, where the atoms Co and Fe occupy the Wyckoff positions 2c and 2d respectively. However, it is now well-established that B-site ordered structure produces (011) Bragg reflection peak in its diffraction pattern refined with $P2_1/n$ symmetry [8]. The absence of such a peak (011) in our experimental pattern at room temperature (300 K) rules out the possibility of B-site ordered structure of PCFO. Subsequent attempt in fitting the data with orthorhombic Pnma space group was successfully done, thus confirming the random distribution of Co and Fe ions. (Fig. 3(a)). Thereby, a disordered orthorhombic Pnma structure has been inferred, where the Co and Fe atoms arbitrarily sit on the crystallographic positions 6c. The calculated structural parameters such as lattice parameters (a,b,c and angles $\alpha, \beta, \gamma$), atomic positions, bond lengths and bond angles are summarized in table 2. It is interesting to note that distortion in the $Co(Fe)O_6$ octahedra is evident from the reduced bond angle of Co(Fe)-O1-Co(Fe), which is found to be $159.87^0$, hence the angle of distortion as obtained from the same formula used in XRD study, is $\delta=10.06^0$. This result again supports the XRD data which also suggested similar octahedral distortions. Again, the theoretical average bond length of Co/Fe-O for $Fe^{3+}$ (HS) and $Co^{3+}$(LS) is 1.97Å (calculations shown in XRD study). Now, from the NPD data analysis, it can be seen from table. 2 that bond lengths Fe/Co-O (2) and Fe/Co-O (1) are 1.974 Å and 1.9517 Å respectively. Thus the bond length Fe/Co-O (2) shows a close match with the theoretical bond length for $Fe^{3+}$ (HS) and $Co^{3+}$(LS) ions. Again, the theoretical average bond length of Fe/Co-O is 2.01 Å for high spin states of both the ions i.e. $Co^{3+}$(HS) and $Fe^{3+}$ (H.S), which does not match with any of the experimentally obtained Fe/Co-O(1 or 2) bond lengths. This is again supporting the low spin state of the $Co^{3+}$ (LS) ions which we predicted earlier from X-ray diffraction and magnetization data analysis.

Fig. 3(d) is showing the intensity of the magnetic super-lattice reflection as a function of temperatures, it can be clearly observed that the intensity shows a drastic jump around 270 K (above which it was almost zero) thus suggesting a second order magnetic phase transition. Interestingly, NPD data recorded at 6 K shows a prominent magnetic super-lattice peak at around ~ $16^0$ which was absent at room temperature (300 K), thus this is a clear and direct evidence of long range magnetic ordering of the $Fe^{3+}$ spins (Fig. 3(b)). Both the 300 K and 6 K data were successfully fitted with Pnma symmetry, thus the observed super lattice reflection peak (011) is not associated to any structural change; rather it is of magnetic origin. The NPD pattern analysis yielded a $G_zF_y$ type of spin ordering which is a canted AFM type of magnetic structure. In this structure, the FM moment is directed along y direction while the G-type of magnetic ordering is occurring along z-direction. This predicts that the system should exhibit dominating antiferromagnetic behaviour. However, the canting of the spins predicts that FM behaviour should co-exist with the dominating AFM background. The microscopic spin arrangements in the $G_zF_y$ magnetic structure is shown by a schematic diagram in the Fig. 3(c). The magnetic moment analysis from NPD pattern gives the moment values 1.9 $\mu_B$ and 0.6 $\mu_B$ from the AFM and FM contributions respectively. Thus, again it suggests dominance of AFM over FM behaviour. As a matter of fact, the total moment calculated for the present system is found to be ~2 $\mu_B$( $\sqrt{[1.9^2 + 0.6^2]} = 2\mu_B$) which is close to the theoretically expected total moment ~2.5 $\mu_B$ for $Co^{3+}$(LS) and $Fe^{3+}$ (HS) ions for the above magnetic structure. In the contrary, for $Co^{3+}$ (HS) and $Fe^{3+}$(HS) ions, the theoretically predicted total moment is 4.52 $\mu_B$ which is much higher value than our experimentally obtained value (2 $\mu_B$). Thus, the analysis confirms the low spin state (LS) for the $Co^{3+}$ ions.

### D. X-ray absorption spectroscopy (XAS) and x-ray magnetic circular dichroism (XMCD) Study

Eventually, a prior understanding of the electronic structures of the constituent elements can essentially help in explaining the origin of different physical properties, especially the magnetic properties. The synchrotron based x-ray absorption spectroscopy (XAS) is a powerful spectroscopic technique to probe the electronic states of a matter. Hence, we have studied the electronic structure of PCFO by employing XAS as well as x-ray magnetic circular dichroism (XMCD) measurements. The XAS spectra have been collected at $L_{2,3}$ edges of Co and Fe by the total electron yield (TEY) mode because of its relatively simple setup and high signal to noise ratio.

Fig. 4(a) demonstrates the Co 2*p* XAS spectrum (at 300 K) related to the photo-absorption from Co2*p* core level to the Co 3*d* unoccupied level. Factually, the Co2p XAS spectra recorded at L*2-3* edge is extremely sensitive to the spin states since it involves the relevant valence shells directly. The spectrum comprises of two main peaks Co$L_3$(2$p_{3/2}$) and Co$L_2$(2$p_{1/2}$) at ~780.7 eV and ~795.2 eV, respectively. The separation of these two peaks is associated to the spin-orbit (SO) coupling (with SO separation energy: ΔE~ 14.5 eV). The line shape and the peak positions of the observed Co2p XAS spectra clearly suggest presence of trivalent Co ions in PCFO [51]. No trace of a pronounced peak at ~ 777 eV corresponding to $Co^{2+}$ ions can be observed from Fig. 4(a)[52]. This directly rules out any possibility of presence of any divalent Co ions in PCFO. The Co*2p* XAS spectrum is a manifestation of the multiplet structure originated from the Co 2p-3d, 3d-3d exchange and Coulomb interactions, as well as fromthe hybridization with the O*2p* ligands and the local crystal field effects [53]. The X-ray absorption, dipole selection rule is capable of precisely estimating the final state (with its relative intensity) $2p^5 3d^{n+1}$ which is to be occupied starting from an initial state $2p^6 3d^n$ (where n=6 for $Co^{3+}$ ion). This is the underlying process which makes the XAS technique to be highly sensitive to the symmetry related to the initial states, e.g. spin states of $Co^{3+}$ ions [53]. This is why several theoretical simulation studies have effectively reproduced the XAS spectra related to different spin states of same magnetic ion. As a matter of fact, on looking at the $L_2$ edge of the Co*2p* XAS spectra, a narrow and relatively sharp peak can be observed which is similar to the feature observed in Co*2p* XAS for $LaCoO_3$ at 20 K [51]. This narrow and sharp $L_2$ peak at Co*2p* XAS is a hallmark for low-spin state (LS) of $Co^{3+}$ ions, thus it undoubtedly confirms the presence of LS $Co^{3+}$ ions in PCFO at room temperature [51]. Thus, XAS data eventually supports the previous XRD and neutron data analysis which also predicted $Co^{3+}$ in LS state. Moreover, inset of Fig. 4(a) shows the XMCD spectra at Co$L_{2,3}$ absorption edge which is calculated by taking difference between XAS spectra under +1 T and -1 T magnetic fields. However, we could not detect any XMCD signals for this case, which clearly suggests that there is no magnetic ordering present due to $Co^{3+}$ ions.

Fig. 4(b) depicts the Fe2*p* XAS spectrum recorded at 300 K. The Fe2*p* XAS spectrum is ascribed to the transition of electrons from Fe2*p* to Fe3*d* states. The Fe 2*p* XAS spectrum can be broadly divided into two peaks Fe$L_3$(2$p_{3/2}$) and Fe$L_2$(2$p_{1/2}$) positioned at ~710.2 eV and~723.6 eV, respectively, the corresponding spin-orbit splitting energy is ΔE~ 13.4 eV. Due to crystal field splitting, each of the main $L_3$ and $L_2$ peaks is further split into $e_g$ and $t_{2g}$ doublet. These $t_{2g}$ features can be observed in the form of a prominent shoulder and a peak just 1.6 eV below the main $L_3$ and $L_2$ peaks respectively. The formation of this $t_{2g}$ and $e_g$ splitting can be attributed to the localized nature of Fe 3*d* electrons. Essentially, the spectral features are similar to the Fe2*p* XAS spectra of the extensively studied $Fe_2O_3$ system, where the nominal valency of the Fe ions is +3 [54]. The Fe2*p* XAS spectral feature excludes similarities from the spectral features as typically seen in metallic Fe, FeO or $Fe_3O_4$, suggesting absence of any mixed-valence states [54]. It can be further noted that for $Fe^{3+}$ ions sitting in the tetrahedral co-ordination, the $L_3$ and $L_2$ peaks are not split into $e_g$-$t_{2g}$ doublet [55]. On the other hand, for the $Fe^{3+}$ ions sitting in the octahedral co-ordination with the oxygen ligands, the $L_3$ and $L_2$ peaks split into two discernible peaks/shoulders namely $e_g$ and $t_{2g}$which are separated by 1.6 eV [55]. The reason behind such differences in the Fe*2p* XAS spectra for different co-ordinations of Fe ions, can be interpreted simply by ligand field theory [55]. It has been consistently shown by ligand field approach that the crystal field splitting is much larger for octahedral co-ordination of Fe with ligands than that for its tetrahedral co-ordination. Thus, by observing the $e_g$ and $t_{2g}$ splitting of the Fe$L_{2,3}$ peaks in the present system PCFO, octahedral co-ordination of $Fe^{3+}$ ions can be confirmed.

Furthermore, the XMCD spectra at Fe$L_{2,3}$ absorption edge which was obtained simply by taking difference between XAS spectra under +1 T and -1 T magnetic fields has been shown in the inset of Fig. 4(b). Even though, the observed XMCD signal is very weak, the signal can be seen (after multiplying it by a factor of 10) in the above figure. The XMCD signal is clearer at $L_3$ edge as compared to that observed at $L_2$ edge. According to the sum-rule, the observation of XMCD signal at the same side (though very weak for $L_2$ edge) suggests that the orbital contribution is dominating in the signal as compared to spin contribution. However, the observation of weak XMCD signal at room temperature is seemingly associated to the presence of short range correlations among the $Fe^{3+}$ spins even above the magnetic transition temperature.

**E. Magnetization Study:**

The temperature (T) variation of magnetization (M) of PCFO sample following the standard zero-field cooled (ZFC) and field cooled (FC) protocols at an applied dc field of 250 Oe, has been illustrated in Fig. 5(a). The magnetization curve displays a sharp jump which is a characteristic of a magnetic transition below $T_N$ ~ 269K, which corresponds to the long range AFM ordering of B-site spins. The exact transition temperature is identified from the inflection point of temperature dependent (dM/dT) curve at 269K, suggesting the long range magnetic ordering (Fig. 5b). To further probe the nature of the magnetic transition, we have recorded the ac susceptibility data around this temperature (Fig. 5 c). The sharp and frequency independent ac $\chi'$ peaks at ~ 269 K confirm the long range magnetic ordering below this transition [15]. Interestingly, at lower temperature~25 K, another relatively broad anomaly is observed in dM/dT, which is an indication of existence of another magnetic phase at low temperatures. The long range ordering is confirmed by the observation of frequency independent sharp peaks of real ac susceptibility $\chi'$ at 269 K [15]. The FC and ZFC arms also show a thermo-magnetic irreversibility or bifurcation below the magnetic ordering temperature $T_N$~269 K, suggesting existence of competition between different magnetic interactions or spin frustrations.

The isothermal field dependent magnetization (M-H) curves at 265 K and 250 K have been recorded to further explore the nature of the magnetic ordering below the magnetic transition temperature $T_N$ ~ 269 K (Fig. 5d). For both of the curves, existence of small hysteresis can be discernible. The exhibition of hysteretic nature with the coercive field of the MH loops is a characteristic of common ferromagnetic (FM) or ferrimagnetic (FIM) materials due to blocking of the domain wall motion. However, no signature of magnetic moment saturation can be seen even at such a high field of 4 KOe, rather it increases monotonically yielding a magnetic moment of $0.12\mu_B$/f.u. (@250 K) thus indicating predominant canted anti-ferromagnetic uncompensated spin ordering in the sample. AFM nature of the sample can be attributed to the anti-parallel alignment of $Fe^{3+}$ spins due to AFM $Fe^{3+}/Fe^{3+}$ interactions. The weak ferromagnetism rises due to the canting of $Fe^{3+}$ spins which can be elucidated by the Dzyaloshinsky-Moriya interaction which is an intrinsic characteristic of canted AFM orthoferrite systems [38-40]. As $Co^{3+}$ ions are in non-magnetic LS state, it does not play any major role in magnetic interactions in the system. To further investigate the type of spin ordering, virgin curves of M-H loops have been plotted as Arrotplot : $M^2$ Vs H/M at temperatures 265 K and 250 K at the inset of Fig. 5(d) [56]. We obtain a negative intercepton the $M^2$ axis by making a linear extrapolation to $H \rightarrow 0$ Oe of the higher field portion of the Arrot plot which confirms the absence of spontaneous magnetization below the transition temperature. This clearly suggests the dominating AFM nature of the sample. Additionally, according to Banerjee's theory, a positive or negative slope of the Arrot curves is indicative of second or first order magnetic phase transition respectively [57]. Hence, the obtained positive slope of the Arrot curves in our case confirms the second order phase transition occurring at 269 K.

Most interestingly, in the "temperature variation of inverse susceptibility $\chi^{-1}$ (T)" curves (Fig. 6) at different applied fields ranging from 250 Oe to 3 T, a rapid down-turn deviation from CW behaviour occurs at temperatures well-above the magnetic ordering temperature ($T_N$ ~269 K). This is a clear indication of nucleation of small but finite sized correlated regions and clusters having short range magnetic ordering embedded in the global paramagnetic matrix above the PM-AFM transition temperature: this is the signature behaviour of Griffiths phase described by Griffiths and Bray's theory, a special and peculiar magnetic phase where the system neither behave like a paramagnet nor shows long range ordering [58-59]. It should be mentioned here that the observation of down-turn behaviour of $\chi^{-1}$ (T) at low fields is very crucial as it eventually helps one to distinguish the Griffiths phase from other non-Griffiths like clustered phases where $\chi^{-1}$ (T) deviates from CW law by showing a up-turn above ordering temperature[60]. From Fig. 6, it is also clear that the down-turn deviation gets softened with increasing magnetic fields and with sufficiently high magnetic fields, yielding to like CW like behaviour, which is also a hallmark for Griffiths phase [60]. It is because of the fact that the magnetization increases linearly with magnetic fields in paramagnetic regions, thus at high fields, PM susceptibility dominates over the contribution from the correlated clusters to the susceptibility. Above the Griffiths phase temperature $T_G$ referring to the highest magnetically ordering temperature, the system enters in purely paramagnetic region. However, in Griffiths phase region, the magnetization fails to behave like an analytical function of magnetic fields. In 1969, in his original pioneering paper, Griffiths theoretically considered a percolation like problem in an Ising ferromagnet, where random dilution has been done by replacing the magnetic ions with non-magnetic ions or simply by creating vacancies [59]. Thus, the nearest neighbour with magnetic ions, the exchange bond strength is J occurring with a distribution probability p while the disorder introduced in the form of non-magnetic ions having bond strength 0 with the corresponding probability (1-p). In this scenario, the co-operative ferromagnetism cannot be established below a critical percolation threshold $p_c$ of the associated lattice, since the theoretical probability for formation of infinite percolating "backbone" is zero (or divergence of correlation length is not possible). In case of p exceeding threshold $p_c$, however, a relatively weak ferromagnetism is established due to shortage of percolation path but certainly at a temperature $T_c^R$ below the long range FM ordering temperature of undiluted system $T_G$ (=$T_c^R$ @ p=1). The Griffiths phase regime is thus defined by the temperature interval of $T_c^R(p)$< T< $T_G$, where singularities occur in the thermodynamic properties (e.g. magnetization) which become non-analytical function of fields, thus the system neither behaves like purely paramagnetic nor can attain long range FM order by forming infinite percolating chain. Later, Bray and Moore generalized this argument for any type of bond-distribution (not only bonds having strengths J and 0) formed due to disorder that eventually reduces the long range magnetic ordering temperature $T_G$ to $T_c$, thus it greatly helped recognizing Griffiths phase in various magnetic systems [58]. Factually, though the experimental realization of Griffiths phase was initially thought to be remote, Salamon et al was first to report an experimental observation of GP by magnetic susceptibility measurements on a hole doped manganite system [61]. In Griffiths phase regime, it doesn't follow CW law rather it follows the power law of inverse susceptibility with a characteristic non-universal exponent λ (positive and lower than unity) describing Griffiths singularity [61-62];

$$\boldsymbol{\chi^{-1}(T) \propto (T-T_c^R)^{1-\lambda}, \ (0<\lambda<1)}$$

Here, it is clear that the aforementioned power law is a modified version of CW law, where the parameter λ is a measure of deviation from CW behaviour. So, to further investigate the result, we have fitted our inverse susceptibility curve at H=250 Oe with above formula. The Griffiths phase temperature is estimated to be $T_G$~ 370 K below of which the sharp down-turn behaviour is observed violating the CW law. Now, in above formula, value of $T_c^R$ is so chosen that the fitting in the paramagnetic region above $T_G$, yielding $\lambda_{PM}$ ~ 0, which is the same procedure as followed by Pramanik et al [63]. The inset (top) of Fig. 6, showing the $\log_{10}$-$\log_{10}$ plot of $\chi^{-1}$ Vs (T-$T_c^R$), where the linear fitting in the Griffiths phase region (T<$T_G$) gave thevalue of $\lambda$ ~ 0.88 which is consistent with the Griffiths phase, thus confirming the existence of Griffiths phase in PCFO. Eventually, the susceptibility in the Griffiths phase region is the manifestation of the sum of two magnetic contributions: paramagnetic susceptibility $\chi_{PM}$ and susceptibility due to magnetically ordered rare region $\chi_R$. When the rare magnetic region (T <$T_G$) is ferromagnetic (FM), for low fields, $\chi_R$ dominates over $\chi_{PM}$, thus results in down-turn behaviour of below $\chi^{-1}$(T) below Griffiths temperature $T_G$. Albeit, if the rare region is anti-ferromagnetic (AFM), the condition of $\chi_R$>$\chi_{PM}$ may not be satisfied, thus the down-turn behaviour which is the hall-mark of Griffiths phase may not be observed. This is the reason why observation of Griffiths phase by susceptibility measurements is extremely rare in AFM systems and it is observed in FM systems mostly. Hence, observation of GP in antiferromagnetic PCFO system is rare as well as very interesting. To date, there are only very few recently reported papers on such observation of Griffiths phase in AFM systems. For example, in the AFM spin chain compound $Ca_3CoMnO_6$, GP was explained through the rise of short range FM correlations due to competing AFM-FM interactions occurring in the ↑↑↓↓ type spin ordering [64]. In another current report on isovalent half doped AFM manganites $R_{0.5}Eu_{0.5}MnO_3$, the presence of GP has been also interpreted based on rise of ab-plane FM superexchange interactions arising due to the structural disorder driven phase inhomogeneity [65]. Another recent report on GP in a geometrically frustrated AFM intermetallic compound $GdFe_{0.17}Sn_2$ , where the observation of GP was again explained by the means of small sized FM clusters driven by the systems inherent non-stoichiometry[66]. In a very recent report, another geometrically frustrated AFM system $DyBaCo_4O_{7+\delta}$ was found to exhibit GP behaviour [67]. The short-range correlations arising due to interactions of $Co^{2+}$/$Co^{3+}$ions sitting in the Kagome and triangular sublattices, seemed to be responsible for the Griffiths singularity in this case.

However, it is expected that spin dynamics in the Griffiths phase region will be different from that in the paramagnetic region. It is because of the fact that the correlated clusters in the Griffiths phase region will relax quite slowly as compared to the spins in PM region. Bray argued that the spin dynamics in the GP region does not follow the exponential decay unlike in the PM region where it obeys the exponential decay. So, he used two models for interacting spins, namely Heisenberg model and Ising model, for investigating the dynamics of the spins in the GP region of diluted magnet [68]. For the diluted rare magnetic region (GP), he defined a spin auto-correlation function C(t) of the form:

$$C(t) \propto \exp\left(-A\left(lnt^{\frac{d}{d-1}}\right)\right) \qquad \text{: For Ising system}$$

$$C(t) \propto \exp(-Bt^{1/2}) \qquad \text{: For Heisenberg System}$$

So, knowing the fact that Griffiths singularities have important effects on the dynamics of the spins, we have carried out isothermal remanent magnetization (IRM) measurements of our sample in the GP regime for further confirmation for the existence of Griffiths phase. The sample was heated upto 400 K with absence of any magnetic fields followed by a cooling to the desired IRM measurement temperatures with applied magnetic field of 1 T. The IRM measurements were done after sudden removal of the magnetic field by measuring the residual magnetization at 300 K and 325 K as a function of time, as shown in the bottom inset of Fig. 6. The time variation of the decay of magnetization for our case, did not fit with exponential power law, thus ruling out the existence of pure PM phase above AFM ordering temperature $T_N$, sub-inset (bottom) of Fig. 6 [69]. However, it is clear from the figure that the IRM curve is best fitted for a decay scheme with spin auto-correlation function C(t) defined for Heisenberg spin model, while it is seen to deviate both from exponential (PM) as well as Ising model decay schemes. Thus, the spin interactions in the current system seem to be following Heisenberg spin model. Again, it suggests the slowing down of the spin dynamics which is expected in a correlated region with short range magnetic ordering. Hence, it is evident of the pre-formation of slowly relaxing clusters with short-range magnetic ordering above long range magnetic ordering temperature $T_N$ ~ 269K, thus elucidating the existence of GP in PCFO system.

However, there are few reports addressing the B-site disorder to be the active source of GP in some perovskite systems, because it introduces inhomogeneous magnetic distribution and drastically reduces the spin/orbital coupling[63,70]. In the pioneering work published by Imry and Ma, the random quenched disorder has been reported to hinder the formation of long range magnetic ordering while favouring the nucleation of correlated clusters [71]. Thereafter, quenched disorder has been remained a key factor for producing GP in many systems [61, 71-75]. Thus, the plausible reasons for the observed Griffiths phase in PCFO may presumably be attributed to the existence of quenched disorder in the form of anti-site disorder in the system as well as small structural distortions present in the system due to smaller size of the Pr ions. The anti-site disorder which is in turn a source of arbitrary distribution of exchange bonds giving rise to competing interactions in the system, is a potential source of Griffiths singularities. As already mentioned, $Co^{3+}$ exists in non-magnetic LS state so it is essentially giving rise to random non-magnetic dilution of $Fe^{3+}$ spins. This is again the perfect platform for percolating correlated clusters in PM matrix, thus triggering the formation of Griffiths phase in the system. Additionally, the spin canting due to DM interactions introduces the competitive AFM/FM interactions which together with the anti-site disorder creates random exchange bonds (not only by values but also by their signs), thus leaving the system frustrated which is also responsible for the formation of correlated clusters in the PM matrix, thus the GP. It is noteworthy to mention here that a very recent report on a very similar double perovskite $Pr_2CrFeO_6$, no magnetic long range ordering (thus Griffiths phase is not relevant here) was found, where both $Cr^{3+}$ and $Fe^{3+}$ are in high spin state [76]. The manifestation of such different behaviours by two such similar systems, undoubtedly suggests that the nonmagnetic LS state of $Co^{3+}$ is playing a vital role in emerging the GP in $Pr_2CoFeO_6$. Though, the anti-site disorder does not cause any structural distortions or the strains in the system due to the equivalent ionic radii of $Co^{3+}$and $Fe^{3+}$ ions. Thus, such structural distortions or strain mediated enhancement of GP can be ruled out. On the contrary, Desisenhofer et al argued that static quench disorder introduced by Jahn-teller (J-T) distortion is responsible for the emergence of GP in $La_{1-x}Sr_xMnO_3$ [72]. However, for the present scenario, the evolution of Griffiths phase is not associated to J-T effect as neither $Fe^{3+}$(HS) nor $Co^{3+}$(LS) exhibit J-T effect. Further, in the contradistinction, Salaman et al have explained the onset of GP in $La_{1-x}Ca_xMnO_3$ ($x \rightarrow 0.3$) due to the bending of Mn-O-Mn bond angle causing alterations in the exchange interactions as a consequence of structural distortion from pure cubic perovskite structure triggered by the smaller size of $Ca^{2+}$ ions [61]. This also seems to be plausible explanation for GP in PCFO. The octahedral distortion of $FeO_6$ octahedra causes the concurrent changes in the $Fe^{3+}$-O-$Fe^{3+}$ exchange interactions which may play an important role in the evolution of GP by aiding the cluster formation in the paramagnetic matrix. Irrespective of all the above situations, the main factor in contributing towards the GP remains the random magnetic ($Fe^{3+}$) dilution by $Co^{3+}$(LS) ions due to ASD which emulates a condition where the exchange bonds with different strengths (J) are randomly distributed. This is the perfect situation for percolating the finite size magnetic clusters above the infinitely long range ordering temperature $T_N$, thus giving rise to the evolution of Griffiths phase [59]. However, for AFM compound PCFO, observation of Griffiths phase is unconventional and thus it can shed new light to the understanding of Griffiths phase in AFM based systems. Thus it requires meticulous study to further explore the underlying physics behind it.

As already mentioned above, the dc ZFC and FC magnetization curves show a sudden slope change at low temperatures (the corresponding broad dip is also observed in dM/dT curve near ~ 25 K) suggesting the presence of a secondary phase at lower temperatures. In contrast to the dc magnetization study, ac susceptibility measurements ($\chi'$&$\chi''$) make it possible to probe the dynamics of the spins thus it has become a powerful tool for investigating glassy spin behaviours [77]. Hence, we have carried out the ac susceptibility measurements in the temperature range 2-75 K at different frequencies. The Fig. 7(a) and its inset are showing the temperature variation of imaginary $\chi''$and real $\chi'$ parts of ac susceptibility data. The curves $\chi'(T)$ show clear anomaly below 40 K and it is becoming more prominent with increasing frequency, thus suggesting a slow dynamic spin relaxation process occurring in this region. The corresponding clear peaks in $\chi''$(T) as expected from Kramers-Kronig relations are observed at ~ 34 K. A common spin-glass feature which can be noted in the $\chi''$(T) peaks is the shift in the peak positions towards higher temperatures with increasing frequencies, indicating slow spin relaxations (Fig. 7(a)) [77-80]. Again, these new low temperature peaks are quite broad extending over a temperature interval of ~70 K unlike the long range ordering peaks which were very sharp having a λ-like cusp [15, 78]. This broad shape of the peaks can also be attributed to the "glassy nature" of the system [78]. Thus, noting all these characteristic features of spin glass, it is comprehensible that the system enters in a re-entrant glasslike state at low temperatures (<40 K). It is best understood based on the existing competing FM and AFM interactions in a system, on lowering the temperature, a special magnetic state is attained where the strengths of the both FM and AFM interactions become equivalent leaving the spins to be frustrated [79,81]. However, the cluster glass state is evolved if one of these competing interactions (AFM or FM) is weaker relative to the other. In cluster glass state, the disorder or spin frustration occurs locally in small region of clusters [80,82-84]. Notwithstanding the complexity, RSG state was nicely described by mean-field model as used by Sherrington- Kirkpatrick for Ising spin systems and the model introduced by Gabay and Toulouse for Heisenberg spin systems [85-86]. According to this model, long range order parameter still remains in the RSG state, briefly which can be described as a state where both the spin-glass state and the long range magnetic correlation co-exist.

To further investigate the spin dynamics and to get the more detailed insight into this RSG state, we have fitted the above data in different models. The frequency dependence of the freezing temperature ( $T_f$) can be calculated by Mydosh parameter (p) which is defined as [78]:

$$\boldsymbol{p} = \frac{\Delta \boldsymbol{T_f}}{\boldsymbol{T_f} \Delta \boldsymbol{log_{10}(f)}},$$

Where $\Delta T_f = T_{f1} - T_{f2}$ and $\Delta log_{10}(f)$= $log_{10}(f1)$- $log_{10}(f2)$. This empirical parameter is a universal tool to distinguish spin-glass state from the super-paramagnetic states [78, 87]. For, typical spin glass or cluster glass systems, the value of p lies between 0.005 and 0.08, while for super paramagnetic system, it is greater than 0.2. The obtained value of p for our experimental graphs of PCFO, is ~ 0.05 which confirms the glass type state. In a spin or cluster glass state, the spin dynamics gets slowed down below the critical temperatures, thus the spins cannot follow the time-varying ac fields and consequently they get frozen randomly. This critical slowing down of spins near the freezing temperatures, can be investigated using the dynamic scaling law [80, 88-89]

$$f = f_0 \left(\frac{T_f - T_{SG}}{T_{SG}}\right)^{z\nu};$$

Where the $f$ is the excitation frequency, $T_{SG}$ is the equivalent spin glass freezing temperature in the limit of $f \rightarrow 0\ Hz$ and $H_{DC} \rightarrow 0\ Oe$, $f_0$ is related to the characteristic spin flipping time ($\tau_0$) as $f_0 = \frac{1}{\tau_0}$ ; zν is the dynamical critical exponent. In Fig. 7(b), "frequency ($f$) Vs freezing temperature ($T_f$)" has been plotted and the best fitting with the above dynamical scaling law yielded: $f_0$ to be 6×10$^6$ Hz ($\tau_0$ = 1.67×10$^{-7}$ s) , $T_{SG}$ = 29 K which is near to the observed spin glass freezing temperatures, the exponent zν is found to be ~ 4.6 which is satisfactory for spin glass state (4 <zν < 12). For a canonical spin glass systemthe microscopic spin flipping time $\tau_0$ typically lies between ~ 10$^{-12}$ to 10$^{-13}$ s which is less than the observed value ~10$^{-7}$ s by few orders. The larger spin flipping time is suggesting the observed transition is due to freezing of finite sized clusters rather than individual spins [80, 88-89]. This is because, the clusters take more time to relax as compared to single spins.For further investigations of inter-cluster interactions, the empirical Vogel-Fulcher law which is the modified version of Arrhenius law, can be employed to fit the above curve "$f$ vs $T_f$". The law being of the form [80, 90]:

$$f = f_0 \exp\left(-\frac{E_A}{K_B(T_f - T_0)}\right);$$

Where $f_0$ is a characteristic frequency,$T_0$ formally known as VF parameter, is a temperature representing the strength of inter-cluster interaction strength and $E_A$ is the activation energy. The Fig. 7(c) is showing the fitted graph using the V-F law. The best fitting yielded $f_0$ ~ 10$^6$ Hz (which is of the same order of characteristic frequency obtained from previous dynamic scale fitting), $T_0$=27.45 K and $E_A/K_B$=37.4 K. The comparable values of $T_0$ and activation energy indicate existence of strong inter-cluster couplings in the system. The obtained large value of $\tau_0 = \frac{1}{f_0}$ is again suggesting the presence of interacting magnetic spin clusters [80, 88-89].

Another experimental realization of slow spin relaxation in the spin glass or cluster glass state is the time evolution of thermo-remanent magnetization (TRM) m(t) below the freezing transition temperature $T_f$. The measurement was carried out following field cooled (FC) protocol. The sample was cooled with an applied magnetic field of 0.1 T down to 25 K (below the freezing temperature) and the time dependent magnetization data was recorded after switching off the magnetic field to zero. The normalized magnetization m(t)=$\left(\frac{M_t}{M_{t=0}}\right)$ has been plotted as a function of time and shown in the Fig. 7(d). The time evolution of the isothermal-remanent magnetizationcan be analyzed using KWW (Kohlrausch Williams Watt) stretched exponential equation as given below [83,91]:

$$m(t) = m_0 - m_g exp\left\{-\left(\frac{t}{\tau}\right)^\beta\right\};$$

Here, $m_0$ is associated to the initial remanent magnetization, $m_g$ is representing the magnetization of glassy component, $\tau$ is the characteristic relaxation time constant and $\beta$ is the shape parameter or stretching exponent. Another power law which also often used for the analysis of time variation of isothermal magnetization is m(t) $\propto t^{\pm\alpha}$[91]. However, we tried to fit our m(t) data with both the above relations but found that the best fitting is obtained with the KWW model, shown in Fig 7(d). The fitting was not satisfactory for the power law ($t^{\pm\alpha}$), and not shown here. The KWW fitting is a powerful technique which is widely used for the investigations of the m(t) data for glassy or disordered systems [91]. For different class of disordered systems, the $\beta$ value lies in between 0 and 1. The obtained $\beta$ value for PCFO is ~ 0.52, thus confirming the existence of glassy state at this temperature (25 K). Hence, all these confirm the system enters in a re-entrant cluster glass state at low temperatures. It is relevant here to note that even below the freezing temperature $T_f$ ~34 K, the long range magnetic ordering still exists which occurs for a re-entrant spin or cluster glass systems. However, co-existence of high temperature long range ordering and low temperature glassy state has been reported in systems such as double perovskite disordered ferromagnet $La_2NiMnO_6$, $Lu_2NiMnO_6$, $La_{1.5}Sr_{0.5}CoMnO_6$, spiral magnet $BiMnFe_2O_6$, $Fe_xMn_{1-x}TiO_3$, ferromagnetic alloy $Ni_{77}Fe_1Mn_{22}$ and antiferromagnet $Fe_{0.55}Mg_{0.45}Cl_2$ etc[12,15,16,92-95]. Our system PCFO contains the two major microscopic ingredients for glassy transitions: (i) Site randomness and (ii) Spin canting driven competing AFM and FM interactions. The random spatial distribution of strongly magnetic ions $Fe^{3+}$ and non-magnetic ions $Co^{3+}$(LS) causes the local environment of the magnetic spins to be inhomogeneous. Thus the anti-site disorder leads to the formation of random exchange bonds causing the spin frustration which at low temperatures ends up in random, non-collinear, frozen states of spins. In pure FM or AFM systems, the domain formation involves microscopic time scales but due to the presence of disorder, it causes pinning of the domain wall which essentially gives rise to metastable states which allow the domain walls to reach one state to other by thermally activated hopping. This process does not allow the system to attain an equilibrium state in the experimental time scale leading to non-equilibrium phases like spin-relaxations, aging effects etc [96]. Again the spin canting is also an important ingredient for glassy sate which can eventually cause formation of finite sized spin clusters where there exist sets of non-collinear ferromagnetically or antiferromagnetically coupled spins; it renders the evolution of glassy state in the system [97]. Consequently, the high temperature ($T_N$ ~ 269 K) long range ordered AFM state gets frustrated due to the increasing competition of AFM and FM interactions with decreasing temperature, leading to the re-entrant glassy state. However, the cluster glass state is achieved because of the fact that the AFM interactions remain still dominating over FM interactions, which is the key ingredient for cluster glass state. To, further explore the origin of the glassy behaviour; we have recorded M-H loops at different temperatures as shown in Fig. 8(a)(1-4). It can be noted that the magnetic hysteresis loop has been enhanced appreciably as temperature is cooled down to 200 K, suggesting effective increase in FM interaction strengths (Fig. 8a(1-2)). However, as temperature is further decreased to 125 K, surprisingly the squareness of the loop (which represents the FM nature) got diminished with a decrease in remanence but increase in coercivity (Fig. 8a(3)). This may be a prior indication that the system is entering in a glassy state. Though, no saturation of magnetization can be seen at any temperatures, implying the dominating AFM nature of the sample. All these facts are confirming the existence of competing AFM and FM interactions which at low temperatures end up in a frozen cluster glass state. However, at 5K, MH loop shows an increased magnetization value (@2T) but it has completely lost its squareness (Fig. 8a(4)). This may also be elucidated by the presence of magnetic rare earth $Pr^{3+}$ ions which triggers the complex and short range $Pr^{3+}$-$Fe^{3+}$ and $Pr^{3+}$-$Pr^{3+}$ interactions which become effective only in low temperatures. In dc ZFC and FC graphs shown in Fig. 5(a), a slope change forming a broad bump below 10K can be noted which is also seemingly related to the $Pr^{3+}$-$Fe^{3+}$ and $Pr^{3+}$-$Pr^{3+}$ interactions. In many double perovskites containing magnetic rare earth ions $R^{3+}$, complex low temperature magnetic behaviours have been attributed to short range R-R and R-B (B= Co, Fe,Mn,Ni etc) interactions[98-99]. Thus it seems to be a plausible origin for the observed low temperature behaviours (<10 K).

Another very unusual and interesting metamagnetic behaviour is observed in field dependent magnetization study of PCFO. Fig. 8(b) is showing the ZFC magnetization curves recorded under different fields. The M(T) curve under a moderate field of 250 Oe increases monotonically with decreasing temperature. To our surprise, for an increased applied magnetic field 600 Oe, M(T) curve shows a dramatic drop and thus forms a peak below the ordering temperature. However, with increasing applied fields (e.g. 1000 Oe etc.) the peak gets flattened and thus becomes broad. However, with application of further higher fields, the peak starts fading away and finally disappears with sufficiently high fields (>1 T). To elucidate this field induced transition, we may consider the strong anisotropy that is present in the system. It seems, the moderate field of 250 Oe was not sufficient for complete anti-parallel alignments of the $Fe^{3+}$ spins due to the strong inherent anisotropic fields. Hence, the monotonous rise in the magnetization with moderate field (250 Oe) is a manifestation of the presence of some uncompensated spins in the system. However, for the intermediate field of 600 Oe, complete anti-parallel alignment of the $Fe^{3+}$ spins is established, thus the magnetization falls drastically resulting in a peak. As the field is further increased, it will try to align the $Fe^{3+}$ spins along the field; hence it will diminish the peak. It is a common feature of AFM systems where application of high fields suppresses the peak intensity [8].

It is now a well-established fact that existence of multiple magnetic phases results in exchange bias effect, a phenomenon where the horizontal or vertical displacement of isothermal magnetization (M) vs field (H) curves occurs [100]. Rigorous theoretical and experimental studies have revealed that exchange anisotropy across the interfaces of different inhomogeneous magnetic phases such as FM/AFM, FM/Spin glass, FM/Ferrimagnet, hard/soft phases of FM systems is responsible for the observation of such exchange bias effect [100-102]. Knowing the fact that the present system PCFO holds multiple magnetic phases including AFM, FM and spin glass at low temperatures, we got motivated to investigate the exchange bias effect in this system. We performed the exchange bias measurements in conventional method i.e. the sample was field cooled with the field of +5 T and -5 T down to 5 K, then field (H) variation of isothermal magnetization (H) data were recorded, (Fig. 8(c)). Clear evidence of exchange bias effect can be observed from the prominent horizontal shift of the M-H loops. However, we also have performed the aforementioned measurements at temperatures higher than cluster glass freezing temperature (~34 K), but no such exchange bias effect was observed (not shown here) in sizeable scales. From this, it can be directly inferred that co-existence of cluster glass andlong range AFM interactions raise the exchange anisotropy at their interfaces, as a consequence exchange bias effect is evolved. To get quantitative value of the exchange bias, we have measured the loop asymmetry along the field and magnetization axes as $H_{CEB} = \frac{[H_{c1} - H_{c2}]}{2}$ and $M_{CEB} = \frac{[M_{r1} - M_{r2}]}{2}$ respectively, where $H_{C1}$ and $H_{C2}$ are the negative or positive intercepts along the field axis of the hysteresis loops recorded with +5 T and -5 T respectively, similarly $M_{r1}$ and $M_{r2}$ are the negative or positive intercepts along the magnetization axis of the said curves. The obtained conventional exchange bias (CEB) values for the current system PCFO are quite high: $H_{CEB}$ ~ 2175 Oe and $M_{CEB}$ ~ 0.033$\mu_B$/f.u.

The pioneering model used by Wang et al for explaining exchange bias of bulk alloy NiMnIn can be helpful for elucidating the observed exchange bias in this case[103]. At low temperatures, the alloy NiMnIn13 enters in a glassy state while the dominant AFM interactions still exists. Thus, the glassy state remains embedded in the AFM matrix at low temperatures, which results in emergence of strong exchange anisotropy at their interfaces, thus raising the exchange bias in the system. Hence, this explanation is plausible for the current scenario because co-existence of cluster glass state in the dominant long range AFM ordering has been already probed at low temperatures.

As the temperature is lowered through the cluster glass freezing temperature ($T_f$ ~ 34 K), the glassy clusters are formed within the long range AFM matrix. So, in this model we can consider an AFM core which is surrounded by frozen cluster glass shell as shown in Fig. 8(d). During the field cooling through the freezing temperature, the spins in the glassy clusters align along the strong applied field thus forming a soft ferromagnetic (SFM) region inside the AFM matrix. After switching off the field to zero, this SFM spins remain aligned ferromagnetically due to the kinetic arrest in the glassy state. Thus, small SFM regions are effectively embedded in the AFM matrix, giving rise to strong unidirectional exchange anisotropy at their interface as shown in Fig. 8(d). These stable SFM regions produce additional remanence and it remains unaltered due to kinetic arrest even upon subsequent field sweep direction while recording the M-H loops. Thus, the M-H loop gets shifted showing the exchange bias effect. Similarly, when the cooling field is applied in the reverse direction, the SFM spins are aligned opposite to the former case and thus producing altered exchange anisotropy at the newly formed interface. Thus, the remanence also gets altered, giving rise to M-H loop shift in the reverse direction. Thus, we can infer that the observed exchange bias is caused by the presence of multiple magnetic phases or inhomogeneous magnetic phases in the system.

## III. Conclusion

Summarizing, we have performed detailed investigation of the magnetic properties of a new member of double perovskite $Pr_2CoFeO_6$ and correlated with its structural and electronic properties. The main interesting aspects of this work which makes $Pr_2CoFeO_6$ as an extremely interesting magnetic system, is the observation of multiple magnetic phases like Griffiths phase, Re-entrant cluster glass (RCG), unusual field induced peak and exchange bias in a single system. The crystal structure investigated by X-ray diffraction and neutron diffraction study highlights the presence of anti-site disorder (ASD) in the B-sites (Co/Fe) and suggests that the system adopts an orthorhombic structure with symmetry Pnma (62). The crucial role played by this ASD in its magnetic properties has been brought out through the neutron diffraction and magnetometry study. A G-type AFM spin ordering has been confirmed from the neutron diffraction study. Ab initio calculations predicted insulating nature of the system. The calculations also showed that anti-ferromagnetic coupling among Fe spins is energetically favourable over the ferromagnetic coupling. Thus, the theoretical calculations agreed well with our experimental observations. Additional information is sought through its electronic structure probed by the XAS study which revealed the charge and spin states of the constituent elements. The nominal valence state of both of the Co and Fe ions found to be +3 while the spin state of the $Co^{3+}$ ions is estimated to be low spin state (LS). The above findings greatly corroborate our arguments that the ASD present in the system is triggered by the same charge states of B-site ions and this ASD along with the LS state of $Co^{3+}$ ions effectively created the random non-magnetic dilution of the magnetic Fe spins. As a matter of fact, the random non-magnetic dilution provides the perfect platform for the preformation of percolating magnetic clusters above $T_N$ ~ 269 K, thus leading to the Griffiths phase in the present system. Again, the exhibition of the Griffiths phase by this antiferromagnetic system $Pr_2CoFeO_6$ is essentially unique since only very few systems e.g. $Ca_3CoMnO_6$, $GdFe_{0.17}Sn_2$, $DyBaCo_4O_{7+\delta}$ are reported so far which order antiferromagnetically and show features of Griffiths phase. Additionally, unique to the current system, it showed Griffiths phase at quite high temperature (269 K$<T_G<$370 K) range. Spin dynamics study by the ac susceptibility study further revealed that the system enters in a RCG state at ~ 34 K where a glassy state is observed to co-exists with global canted-antiferromagnetic state. Again ASD along with the spin canting driven spin frustration played the major role in bringing out this glassiness in the system. The observed low temperature exchange bias (@5 K) is elucidated through the co-existence of AFM and the glassy states and explained through the AFM core and glassy shell model. The results of the present work can significantly provoke the experimental as well as the theoretical investigations to study the possible impact of such anti-site disorder and the spin states on the magnetic properties of different magnetic systems.

## Acknowledgements

The authors thank to the Central Instrumentation Facility Centre, Indian Institute of Technology (BHU) for providing the low temperature magnetic measurementsfacility.

**Table1**: Structural parameters and crystallographic sites determined from Rietveld profile refinement of the powder XRD patterns for **$Pr_2CoFeO_6$** at 300K (room temperature).

Space group: Pnma

| Parameters | Value |
|---|---|
| **Lattice constant (Å)** | **a= 5.4351** |
| | **b=7.6757** |
| | **c=5.4376** |
| | **α=β=γ=90.000** |
| **Cell volume (Å$^3$)** | **226.8454** |
| **Pr site** | **4c** |
| **x** | **0.03203** |
| **y** | **0.2500** |
| **z** | **0.01086** |
| **Co site** | **4b** |
| **x** | **0.00** |
| **y** | **0.00** |
| **z** | **0.50** |
| **Fe site** | **4b** |
| **x** | **0.00** |
| **y** | **0.00** |
| **z** | **0.50** |
| **O(1) site** | **4c** |
| **x** | **0.49054** |
| **y** | **0.25000** |
| **z** | **0.00000** |
| **O(2) site** | **8d** |
| **x** | **0.32906** |
| **y** | **-0.04394** |
| **z** | **0.31110** |
| **$R_{wp}$** | **18.1** |
| **$R_{exp}$** | **14.26** |
| **$R_{wp}/R_{exp}$** | **1.269** |
| **$Chi^2$** | **1.68** |
| **$d_{Pr-Pr}$(Å)** | **3.85541** |
| **$d_{Pr-O(1)}$(Å)** | **2.49276** |
| **$d_{Pr-O(2)}$ (Å)** | **3.06890** |
| **$d_{Co-O(1)}$ (Å)** | **1.91961** |
| **$d_{Co-O(2)}$ (Å)** | **1.95922** |
| **$d_{Fe-O(1)}$ (Å)** | **1.91961** |
| **$d_{Fe-O(2)}$ (Å)** | **1.95922** |
| | **-** |
| **<(Pr)-(O1)-(Pr)>(deg)** | **177.16759 -** |
| **<(Pr)-(O2)-(Pr)>(deg)** | **84.01257** |
| **<(Fe)-(O1)-(Fe)>(deg)** | **164.5754** |
| **<(Co)-(O1)-(Co)>(deg)** | **164.5754** |
| **<(Fe)-(O2)-(Fe)> (deg)** | **169.69672** |
| **<(Co)-(O2)-(Co)> (deg)** | **169.69672** |

**Table 2**: Structural parameters and crystallographic sites determined from Rietveld profile refinement of the powder neutron diffraction patterns for **$Pr_2CoFeO_6$** at 300K and 6 K.

Space group: Pnma

| NPD data recorded at | 300 K | 6 K |
|---|---|---|
| Lattice constant (Å) | a= 5.44592 | a= 5.45221 |
| | b= 7.68672 | b= 7.67367 |
| | c= 5.43325 | c= 5.42731 |
| | α=β=γ=90.000 | α=β=γ=90.000 |
| Cell volume (Å$^3$) | 227.4427 | 227.0703 |
| Pr site | 4c | 4c |
| x | -0.0334(5) | -0.0358(9) |
| y | 0.25000 | 0.25000 |
| z | -0.0109(11) | -0.0144(17) |
| Co site | 4b | 4b |
| x | 0.00000 | 0.00000 |
| y | 0.00000 | 0.00000 |
| z | 0.50000 | 0.50000 |
| Fe site | 4b | 4b |
| x | 0.00000 | 0.00000 |
| y | 0.00000 | 0.00000 |
| z | 0.50000 | 0.50000 |
| O(1) site | 4c | 4c |
| x | 0.5121(6) | 0.5151(9) |
| y | 0.25000 | 0.25000 |
| z | 0.0616(8) | 0.0541(10) |
| O(2) site | 8d | 8d |
| x | 0.2868(4) | 0.2905(6) |
| y | -0.0417(3) | -0.0458(5) |
| z | 0.2859(4) | 0.2855(6) |
| $d_{Pr-O(1)}$(Å) | 2.506(4) | 2.477(7) |
| $d_{Pr-O(2)}$ (Å) | 2.735(4) | 2.755(6) |
| $d_{Co-O(1)}$ (Å) | 1.9517(8) | 1.9425(8) |
| $d_{Co-O(2)}$ (Å) | 1.974(2) | 1.997(3) |
| $d_{Fe-O(1)}$ (Å) | 1.9517(8) | 1.9425(8) |
| $d_{Fe-O(2)}$ (Å) | 1.974(2) | 1.997(3) |
| <(Pr)-(O1)-(Pr)>(deg) | 163.4(2) | 168.2(6) |
| <(Pr)-(O2)-(Pr)>(deg) | 98.8(3) | 98.43(20) |
| <(Fe)-(O1)-(Fe)>(deg) | 159.87(3) | 161.94(4) |
| <(Co)-(O1)-(Co)>(deg) | 159.87(3) | 161.94(4) |
| <(Fe)-(O2)-(Fe)> (deg) | 155.06(9) | 153.26(13) |
| <(Co)-(O2)-(Co)> (deg) | 155.06(9) | 153.26(13) |

References:

1. J. Wang, J. B. Neaton, H. Zheng, V. Nagarajan, S. B. Ogale, B. Liu, D.Viehland, V. Vaithyanathan, D. G. Schlom, U. V. Waghmare, N. A. Spaldin, K. M. Rabe, M.Wuttig, and R. Ramesh, Science **299**, 1719 (2003).
2. S.W. Cheong and M. Mostovoy, Nat. Mater. **6**, 13 (2007).

3. Y. Kitagawa, Y. Hiraoka, T. Honda, T. Ishikura, H. Nakamura, and T. Kimura, Nat. Mater. **9**, 797 (2010).
4. M. Fiebig, T. Lottermoser, D. Fr€ohlich, A. V. Goltsev, and R. V. Pisarev, Nature **419**, 818 (2002).
5. F. S. Galasso, Structure, Properties and Preparation of Perovskite-type Compounds (Pergamon, London, 1969)
6. Y. Kitagawa, Y. Hiraoka, T. Honda, T. Ishikura, H. Nakamura, and T. Kimura, Nat. Mater. **9**, 797 (2010).
7. Y. Guo, L. Shi, S. Zhou, J. Zhao, and W. Liu, App. Phys. Lett. **102**, 222401 (2013).
8. G.R. Haripriya, H.S. Nair, R. Pradheesh, S. Rayaprol, V. Siruguri, D. Singh, R. Venkatesh, V. Ganesan, K. Sethupathi, and V. Sankaranarayanan, J. Phys.: Condens. Matter. **29**, 475804 (2017).
9. L. Balcells, J. Navarro, M. Bibes, A. Roig, Martínez B., and J. Fontcuberta, App. Phys. Lett. **78**, 781 (2001).
10. M.H. Phana, S.B. Tiana , D.Q. Hoanga , S.C. Yua,, C. Nguyenb , A.N. Ulyanov, J. Magn. Magn, Mater. **309**, 258-259 (2003)
11. N.S. Rogado, J. Li, A.W. Sleight, and M.A. Subramanian, Advanced Materials **17**, 2225 (2005).
12. S. Yáñez-Vilar, E.D. Mun, V.S. Zapf, B.G. Ueland, J.S. Gardner, J.D. Thompson, J. Singleton, M. Sánchez-Andújar, J. Mira, N. Biskup, M.A. Señarís-Rodríguez, and C.D. Batista, Phys. Rev. B **84**, (2011).
13. H.S. Nair, R. Pradheesh, Y. Xiao, D. Cherian, S. Elizabeth, T. Hansen, T. Chatterji, and T. Brückel, J. App. Phys. **116**, 123907 (2014).
14. J. Blasco, J. García, G. Subías, J. Stankiewicz, J.A. Rodríguez-Velamazán, C. Ritter, J.L. García-Muñoz, and F. Fauth, Phys. Rev. B **93**, (2016).
15. D. Choudhury, P. Mandal, R. Mathieu, A. Hazarika, S. Rajan, A. Sundaresan, U.V. Waghmare, R. Knut, O. Karis, P. Nordblad, and D. D. Sarma , Phys. Rev. Lett **108,** 127201 (2012)
16. J. Krishna Murthy and A. Venimadhav, App. Phys. Lett. **103**, 252410 (2013)
17. H.S. Nair, D. Swain, H. N., S. Adiga, C. Narayana, and S. Elzabeth, J. App. Phys. **110**, 123919 (2011).
18. T. Chakraborty, H. S. Nair, H. Nhalil, K. R. Kumar, A. M. Strydom and S. Elizabeth, J. Phys. Cond. Mater. **29**, 025804 (2017).
19. R. I. Dass and J. B. Goodenough, Phys. Rev B **67**, 014401 (2003)
20. J. B. Goodenough, Phys. Rev. 100, 564 (1955).
21. M. García-Hernández, J.L. Martínez, M.J. Martínez-Lope, M.T. Casais, and J.A. Alonso, Phys. Rev. Lett. **86**, 2443 (2001).
22. A. S. Ogale, S. B. Ogale, R. Ramesh, and T. Venkatesan, App. Phys. Lett. **75**, 537 (1999).
23. G. Sharma, J. Saha, S. D. Kaushik, V. Siruguri, and S. Patnaik, App. Phys. Lett. **103**, 012903 (2013).
24. K. Yoshimatsu, K. Nogami, K. Watarai, K. Horiba, H. Kumigashira, O. Sakata, T. Oshima, and A. Ohtomo, Phys. Rev. B **91**, 054421 (2015).
25. M. P. Singh, K. D. Truong, S. Jandl, and P. Fournier, J. App. Phys. **107**, 09D917 (2010).
26. J.C. Walling and R.L. White, Phys. Rev. B **10**, 4748 (1974).
27. D. Meier, H. Ryll, K. Kiefer, B. Klemke, J.-U. Hoffmann, R. Ramesh, and M. Fiebig, Phys. Rev. B **86**, 184415 (2012).
28. J. Hemberger, S. Lobina, H.A. Krug von Nidda, N. Tristan, V.Y. Ivanov, A. A. Mukhin, A.M. Balbashov, A. Loidl, Phys. Rev. B **70**, 024414 (2004).

29. J. K. Murthy, A. Venimadhav, J. All. Comp. **719**, 341 (2017).
30. A. V. Kimel, A. Kirilyuk, A. Tsvetkov, R. V. Pisarev, Th. Rasing, Nature **429**, 850 (2004).
31. T. yamaguchi, J. Phys. Chem. Solids. **35**, 479 (1974).
32. M. R. Moldover, G. gjolander, and W. Weyhmann, Phys. Rev. Lett. **26**, 1257 (1971).
33. W. Zhao, S. Cao, R. Huang, Y. Cao, K. Xu, B. Kang, J. Zhang, and W. Ren, Phys. Rev. B **91**, 104425 (2015).
34. H. Wu, S. Cao, M. Liu, Y. Cao, B. Kang, J. Zhang, and W. Ren, Phys. Rev. B **90**, 144415 (2014).
35. I. Dzyaloshinsky, J. Phys. Chem. Solids **4**, 241 (1958).
36. T. Moriya, Phys. Rev. **120**, 91 (1960).
37. D. Treves, Phys. Rev. **125**, 1843 (1962).
38. M.W. Haverkort, Z. Hu, J. C. Cezar, T. Burnus, H. Hartmann, M. Reuther, C. Zobel, T. Lorenz, A. Tanaka, N. B. Brookes, H. H. Hsieh, H.-J. Lin, C. T. Chen, and L. H. Tjeng, Phys. Rev. Lett **97**, 176405 (2006).
39. F. Guillou, Q. Zhang, Z. Hu, C. Y. Kuo, Y. Y. Chin, H. J. Lin, C. T. Chen, A.Tanaka, L. H. Tjeng, V. Hardy, Phys. Rev. B. **87**, 115114 (2013).
40. M. A. Korotin, S. Yu. Ezhov, I. V. Solovyev, and V. I. Anisimov, Phys. Rev. B **54**, 5309 (1996).
41. C. Zobel, M. Kriener, D. Bruns, J. Baier, M. Gru¨ninger, and T. Lorenz, Phys. Rev. B **66**, 020402 (R) (2002).
42. S. Tsubouchi, T. Kyômen, M. Itoh, P. Ganguly, M. Oguni, Y. Shimojo, Y. Morii, and Y. Ishii, Phys. Rev. B **66**, 052418 (2002).
43. V.V. Mehta, S. Bose, J.M. Iwata-Harms, E. Arenholz, C. Leighton, and Y. Suzuki, Phys. Rev. B **87**, 020405 (R) (2013).
44. M. Anderson, K. Greenwood, G. Taylor, and K. Poeppelmeier, Progress in Solid State Chemistry **22**, 197 (1993).
45. A.M. Glazer, Acta Crystallographica Section A **31**, 756 (1975).
46. A. Mu Oz, J.A. Alonso, M.T. Casais, M.J. Mart Nez-Lope, and M.T. Fern N. D Az, J. Phys.: Condens. Matter. **14**, 8817 (2002).
47. R.D. Shannon, Acta Crystallogr., Sect. A **32**, 751 (1976).
48. S. K. Pandey, S. Patil, V. R. R. Medicherla, R. S. Singh and K. Maiti, Phys. Rev. B **77** , 115137 (2008)
49. H. Hsu, P. Blaha, R. M. Wentzcovitch and C. Leighton, Phys. Rev. B **82**, 100406 R (2010)
50. Y. L. Lee, J. Kleis, J. Rossmeisl and D. Morgan, Phys. Rev B **80**, 224101 (2009)
51. M.W. Haverkort, Z. Hu, J.C. Cezar, T. Burnus, H. Hartmann, M. Reuther, C. Zobel, T. Lorenz, A. Tanaka, N.B. Brookes, H.H. Hsieh, H.-J. Lin, C.T. Chen, and L.H. Tjeng, Phys. Rev. Lett. **97**, 176405 (2006).
52. S.I. Csiszar, M.W. Haverkort, Z. Hu, A. Tanaka, H.H. Hsieh, H.-J. Lin, C.T. Chen, T. Hibma, and L.H. Tjeng, Phys. Rev. Lett. **95**, 187205 (2005).
53. Z. Hu, H. Wu, M.W. Haverkort, H.H. Hsieh, H.-J. Lin, T. Lorenz, J. Baier, A. Reichl, I. Bonn, C. Felser, A. Tanaka, C.T. Chen, and L.H. Tjeng, Phys. Rev. Lett. 92, 207402 (2004).
54. D. H. Kim, H. J. Lee, G. Kim, Y. S. Koo, J. H. Jung, H. J. Shin, J.-Y. Kim, and J.-S. Kang , Phys. Rev. B 79, 033402 (2009).
55. K.M. Krishnan, Ultramicroscopy 32, 309 (1990).
56. A. Arrott, Phys. Rev. **108**, 1394 (1957).
57. B.K. Banerjee, Phys. Lett. **12**, 16 (1964)
58. A. J. Bray, M. A. Moore, J. Phys. C **15**, L765 (1982).
59. R. B. Griffiths, Phys. Rev. Lett. **23**, 17 (1969)

60. C. He, M. A. Torija, J. Wu, J. W. Lynn, H. Zheng, J. F. Mitchell, C. Leighton, Phys. Rev. B **76**, 014401 (2007).
61. M. B. Salamon, P. Lin, and S. H. Chun, Phys. Rev. Lett. **88**, 197203 (2002).
62. A.H.C. Neto, G. Castilla, and B.A. Jones, Phys. Rev. Lett. **81**, 3531 (1998)
63. A. K. Pramanik and A. Banerjee, Phys. Rev. B **81**, 024431 (2010)
64. Z.W. Ouyang, N.M. Xia, Y.Y. Wu, S.S. Sheng, J. Chen, Z.C. Xia, L. Li, and G.H. Rao, Phys. Rev. B **84**, 054435 (2011).
65. A Karmakar, S Majumdar, S Kundu, T K Nath and S Giri, J. Phys.: Condens. Matter **25**, 066006 (2013).
66. K. Ghosh, C. Mazumdar, R. Ranganathan, S. Mukherjee, Sci. Rep. **5**, 15801 (2015).
67. J. Kumar, S. N. Panja, S. Dengre, S. Nair, Phys. Rev. B **95**, 054401 (2017)
68. A. J. Bray, Phys. Rev. Lett. **60**, 720 (1988).
69. M. Randeria, J. P. Sethna, and R. G. Palmer, Phys. Rev. Lett. **54**, 1321 (1985).
70. J. Fan, L. Pi, Y. He, L. Ling, J. Dai, and Y. Zhang, J. App. Phys. **101**, 123910 (2007).
71. Y. Imry and S. K. Ma, Phys. Rev. Lett. **35**, 1399 (1975).
72. J. Deisenhofer, D. Braak, H.-A. Krug von Nidda, J. Hemberger,R. M. Eremina, V. A. Ivanshin, A. M. Balbashov, G. Jug, A.Loidl, T. Kimura, and Y. Tokura, Phys. Rev. Lett. **95**, 257202 (2005).
73. Y. Shimada, S. Miyasaka, R. Kumai, and Y. Tokura, Phys. Rev. B **73**, 134424 (2006).
74. S. Guo, D. P. Young, R. T. Macaluso, D. A. Browne, N. L. Henderson, J. Y. Chan, L. L. Henry, and J. F. DiTusa, Phys. Rev.Lett. **100**, 017209 (2008).
75. C. Magen, P. A. Algarabel, L. Morellon, J. P. Ara´ujo, C. Ritter,M. R. Ibarra, A. M. Pereira, and J. B. Sousa, Phys. Rev. Lett. 96,167201 (2006).
76. N. Das, S. Singh, A. G. Joshi, M. Thirumal, V. R. Reddy, L. C. Gupta, A. K. Ganguli, Inorg. Chem. **56**, 12712 (2017).
77. J. Mydosh, J. Magn. Magn. Mater. **157-158**, 606 (1996).
78. J. A. Mydosh, Spin Glasses: An Experimental Introduction (London: Taylor and Francis), 1933
79. S. Niidera, S. Abiko, and F. Matsubara, Phys. Rev. B **72**, 214402 (2005).
80. M. D. Mukadam, S. M. Yusuf, P. Sharma, S. K. Kulshreshtha, and G. K. Dey, Phys. Rev. B **72**, 174408 (2005).
81. K. Jonason, J. Mattsson, and P. Nordblad, Phys. Rev. Lett. **77**, 2562 (1996).
82. S. Mukherjee and R. Ranganathan, P. S. Anilkumar and P. A. Joy, Phys. Rev. B 54, 9267 (1996).
83. R.S. Freitas, L. Ghivelder, F. Damay, F. Dias, and L.F. Cohen, Phys. Rev. B 64, 144404 (2001).
84. W. Kleemann, V. V. Shvartsman, P. Borisov, A. Kania, Phys. Rev. Lett. 105, 257202 (2010).
85. D. Sherrington and S. Kirkpatrick, Phys. Rev. Lett. 32, 1782 (1975).
86. M. Gabay and G. Toulouse, Phys. Rev. Lett. 47, 201 (1981).
87. F. Wang, J. Zhang, Y.F. Chen, G. J. Wang, J.R. Sun, S.Y. Zhang, and B. G. Shen, Phys. Rev. B **69**, 094424 (2004).
88. C. Djurberg, P. Svedlindh, P. Nordblad, M. F. Hansen, F. Bødker, and S. Mørup, Phys. Rev. Lett. 79, 5154 (1997).
89. D. N. H. Nam, R. Mathieu, P. Nordblad, N. V. Khiem, and N. X. Phuc, Phys. Rev. B 62, 8989 (2000).
90. J. L. Tholence, Solid State Commun. **35**, 113 (1980).
91. A. Ito, H. Aruga, E. Torikai, M. Kikuchi, Y. Syono, and H. Takei, Phys. Rev. Lett. **57**, 483 (1986).

92. S. Ghara, B. G. Jeon, K. Yoo, K. H. Kim, and A. Sundaresan, Phys. Rev. B **90**, 024413 (2014)
93. H. Yoshizawa, S. Mitsuda, H. Aruga, and A. Ito, Phys. Rev. Lett. 59, 2364 (1987).
94. P. Z. Wong, S.V. Molnar, T.T.M. Palstra, J.A. Mydosh, H. Yoshizawa, S.M. Shapiro, and A. Ito, Phys. Rev. Lett. 55, 2043 (1985).
95. T. Sato, T. Ando, T. Ogawa, S. Morimoto, and A. Ito, Phys. Rev. B **64**, 184432 (2001).
96. E. Vincent, V. Dupuis, M. Alba, J. Hammann, and J. P. Bouchaud, Europhys. Lett. **50**, 674 (2000).
97. P. Mahadevan, F. Aryasetiawan, A. Janotti, and T. Sasaki, Phys. Rev. B **80**, 035106 (2009).
98. J. Hemberger, S. Lobina, H.-A.K.V. Nidda, N. Tristan, V.Y. Ivanov, A.A. Mukhin, A.M. Balbashov, and A. Loidl, Phys. Rev. B **70**, 024414 (2004).
99. H. Wu, S. Cao, M. Liu, Y. Cao, B. Kang, J. Zhang, and W. Ren, Phys. Rev. B **90**, 144415 (2014).
100. W. H. Meiklejohn and C. P. Bean, Phys. Rev. 102, 1413 (1956).
101. M. Ali, P. Adie, C.H. Marrows, D. Greig, B.J. Hickey, and R.L. Stamps, Nat. Mater. 6, **70** (2006).
102. Q. K. Ong, A. Wei, and X.M. Lin, Phys. Rev. B **80**, 134418 (2009).
103. B.M. Wang, Y. Liu, P. Ren, B. Xia, K.B. Ruan, J.B. Yi, J. Ding, X.G. Li, and L. Wang, Phys. Rev. Lett. **106**, 077203 (2011).

Figure captions:

Figure. 1 : X-ray diffraction pattern along with its Rietveld refinement at 300 K. Inset: Polyhedral representation of the crystal structure. The green and red balls are representing the Pr and O atoms. The blue octahedra refer to $Co/FeO_6$.

Figure. 2 : (a) shows the total density of states (TDOS) as a function of energy (scaled with Fermi energy) for PCFO with AFM coupling in Fe spins. (b), (c), (d) and (e) are depicting the spin integrated partial density of states (PDOS) of Pr (s,p,d and f), Co (s,p, and d), Fe (s, p and d) and O (s and p) for PCFO respectively. (f) Shows the spin resolved PDOS for Pr-f, Co-d, Fe-d, O-p orbitals.

Figure. 3 : (a) and (b) show the powder neutron diffraction (NPD) data @300 K and @6 K with its Rietveld refinements respectively. (c) Depicts the spin ordering obtained from NPD data. (d): Temperature variation of the magnetic reflection as obtained through NPD study.

Figure. 4: (a) and (b) show the room temperature XAS spectra at Co*L2-3* and FeL*2-3* edges respectively. The inset of Fig. (a) and (b) are showing the XMCD data @300 K for Co*L2-3* and FeL*2-3* edges respectively.

Figure. 5: (a): ZFC and FC M(T) curves recorded at H=250 Oe. (b) shows the “dM/dT Vs T” plot @H=250 Oe. (c): Temperature variation of ac $\chi'$ (real) at different frequencies. (d): M(H) curves recorded at 265 K and 250 K. The inset top and bottom show the “Arrot plot” of the M(H) curves at 250 K and 265 K respectively.

Figure. 6: “Inverse susceptibility Vs Temperature” plot at different magnetic fields (H) has been shown to study the Griffiths phase. Inset top is showing the “$\log_{10}$-$\log_{10}$” plot of “$\chi^{-1}$ Vs (T-$T_C^R$)”, where the linear fitting is done to confirm Griffiths phase. Inset bottom shows the “time-dependent thermo-remanent magnetization (TRM)” study at 300 K and 325 K and its Heisenberg fit. The inset shows the Heisenberg, Ising and exponential fit of TRM data at 300 K.

Figure. 7: (a): "$\chi''$ Vs temperature" curves at different frequencies are shown. Inset showing the corresponding $\chi'$(T) curves. Fig. (b) and (c) are showing the dynamic fit and Vogel-Fulcher fit of the $T_f$ (T) data. (d): KWW stretched exponential equation fit of the time evolution of the isothermal-remanent magnetization at 25 K.

Figure. 8: (a) M(H) curves recorded at different temperatures. (b): Field dependent ZFC M(T) curves. (c) M(H) curves at 5 K recorded after field cooling under H=+/-5 T. (d): Simplified schematic picture of the "Core-Shell" model depicting the AFM core being surrounded by CG shell and the consequent rise of the conventional exchange bias.

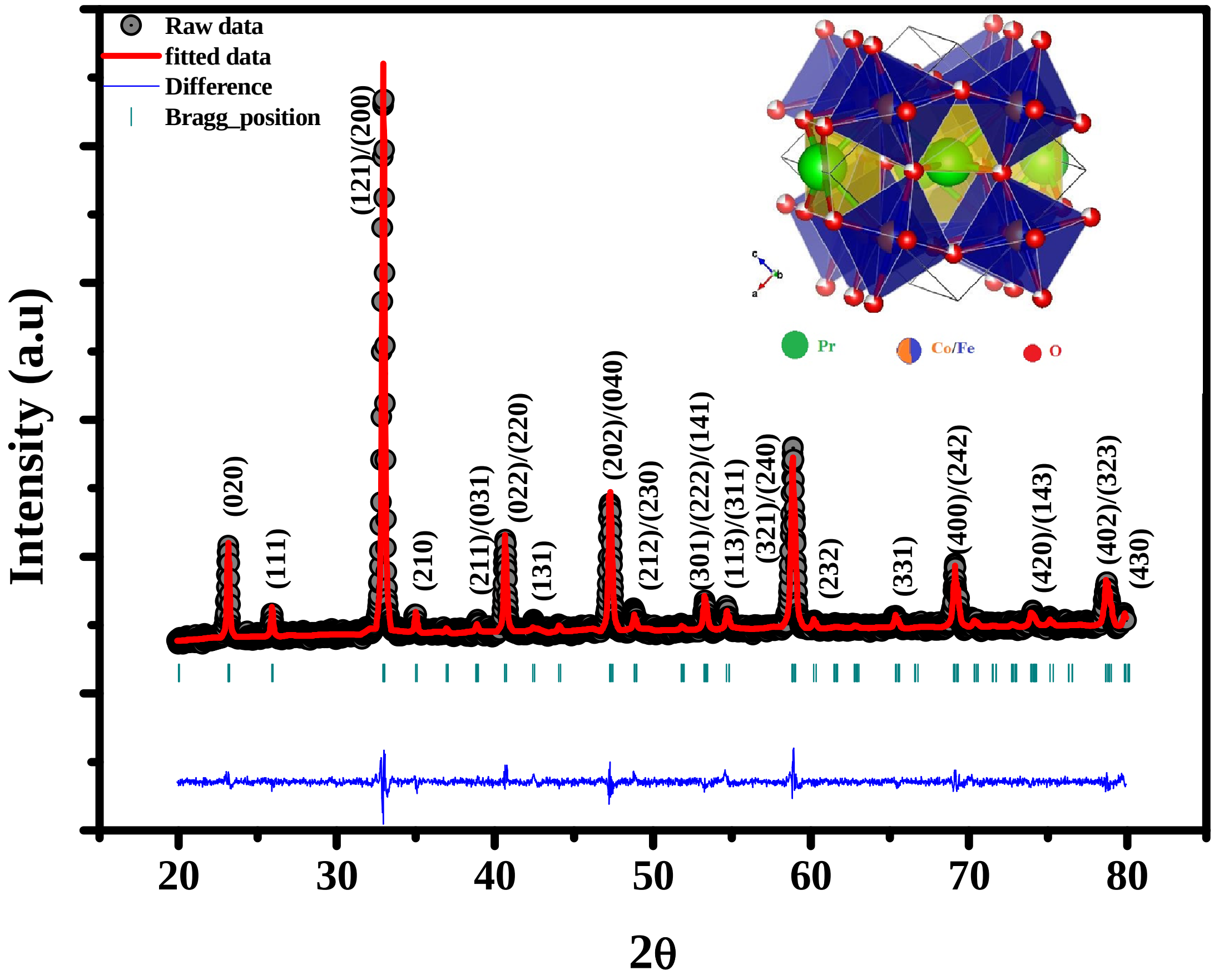

**Figure. 1**

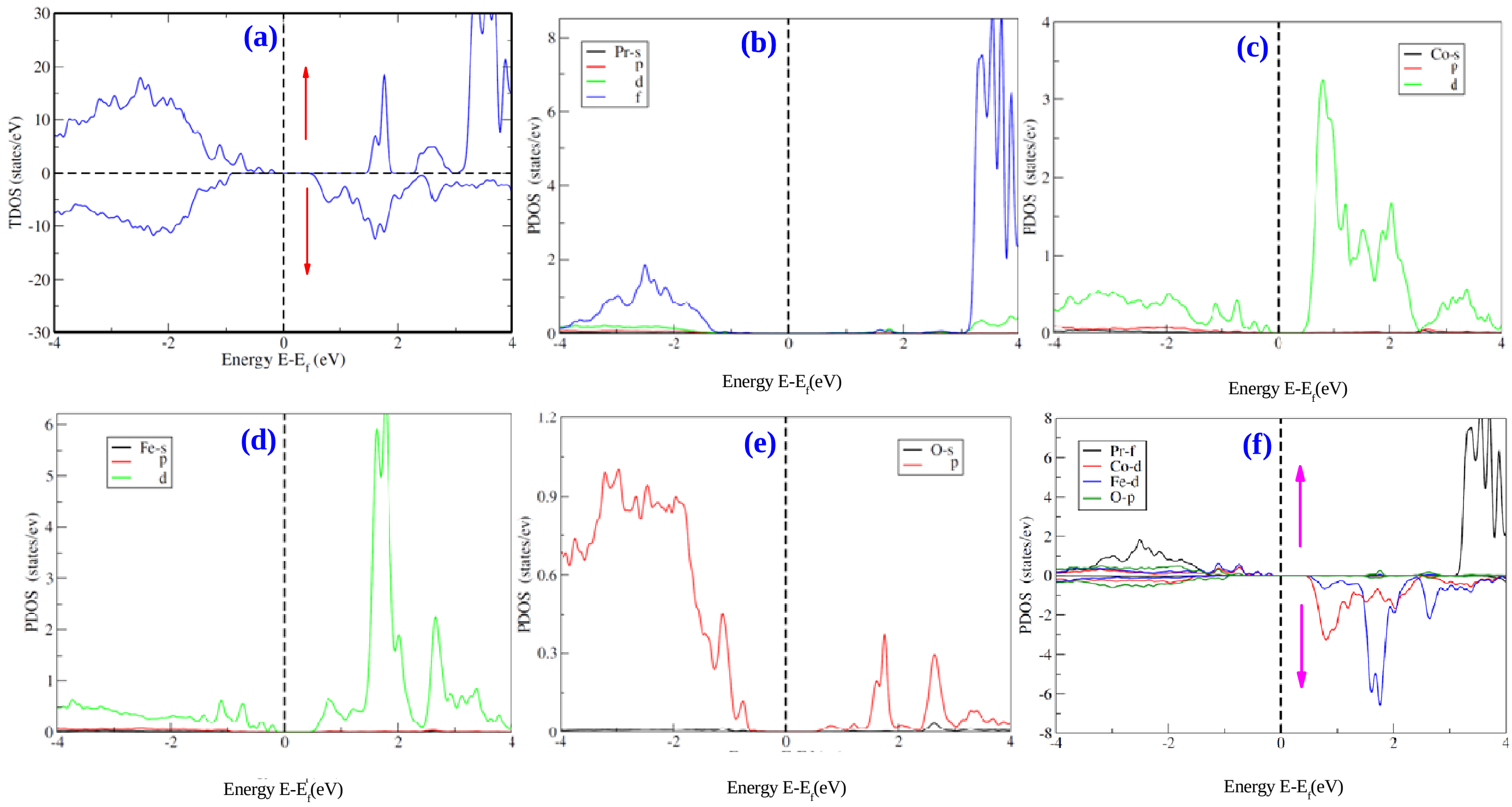

**Figure. 2**

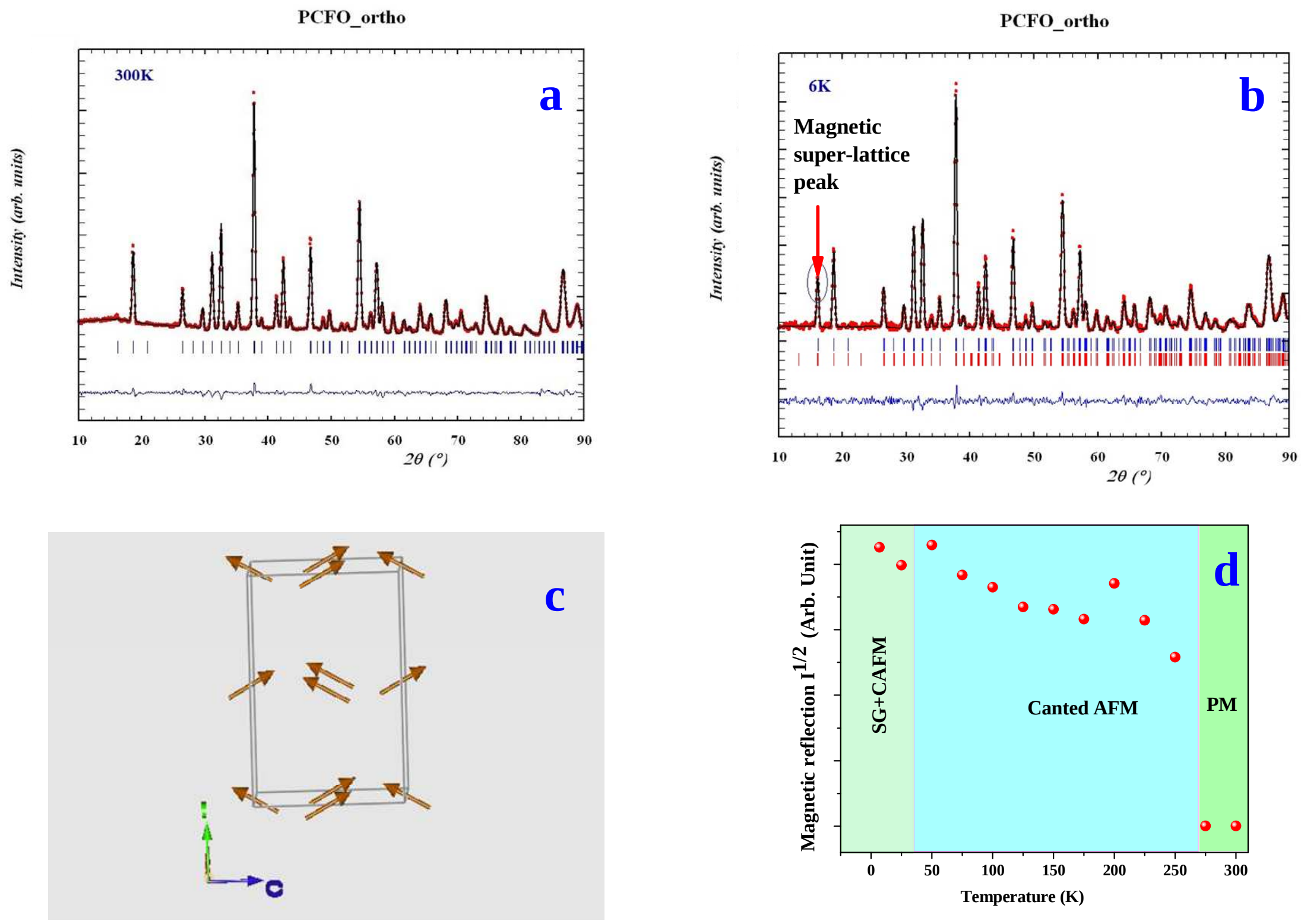

**Figure. 3**

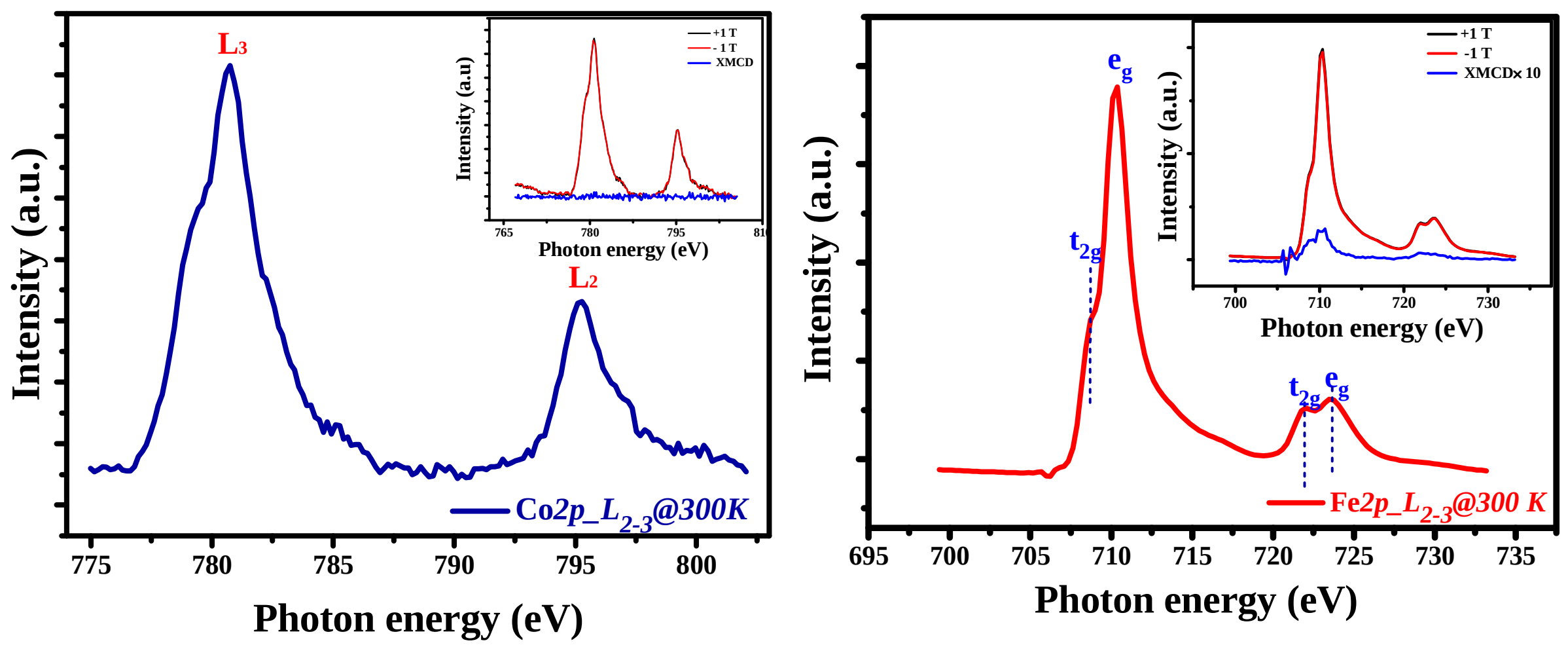

Figure. 4

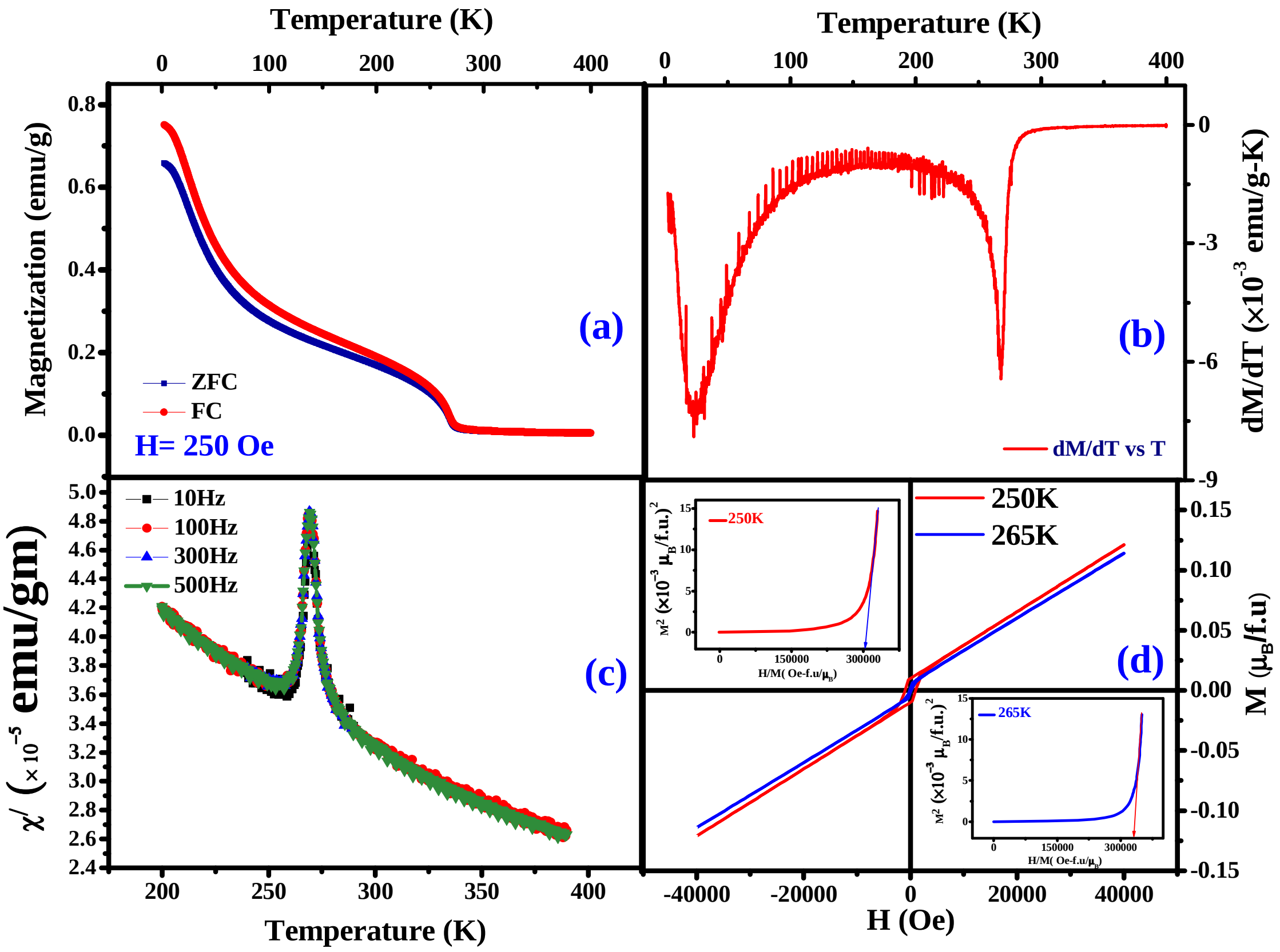

Figure. 5

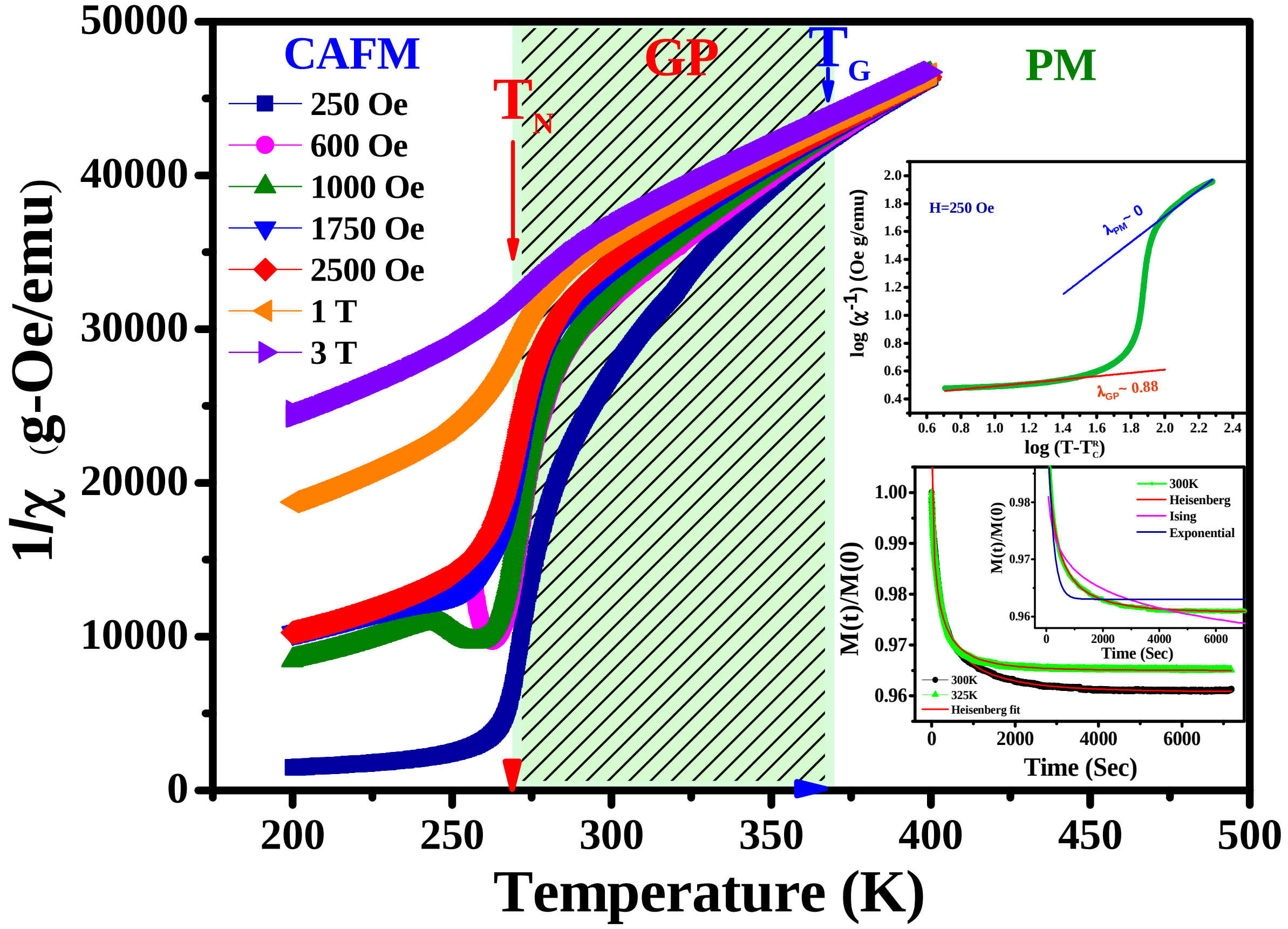

Figure. 6

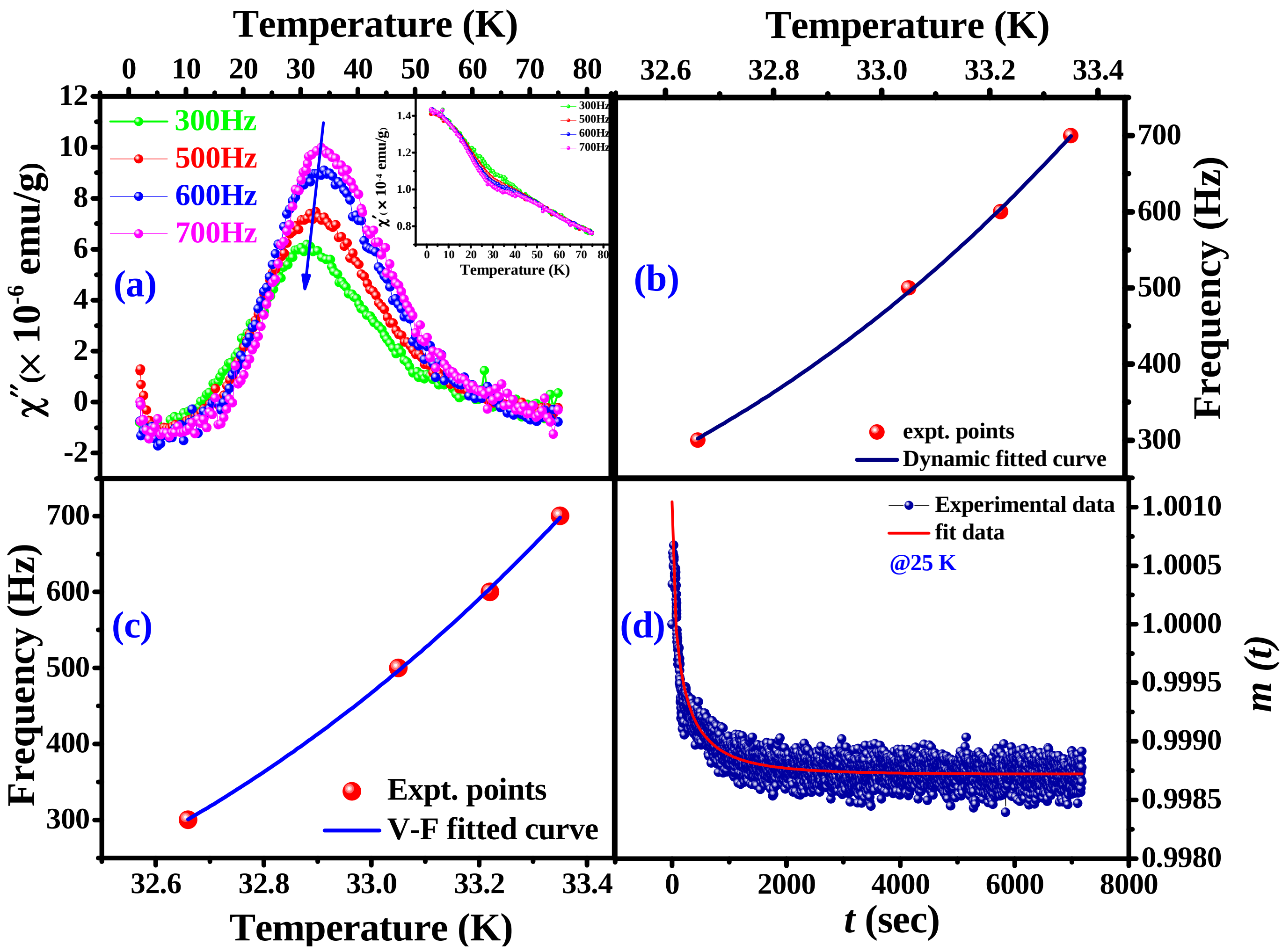

Temperature (K)
0 10 20 30 40 50 60 70 80
300Hz
500Hz
600Hz
700Hz
(a)
χ''(× 10⁻⁶ emu/g)
12 10 8 6 4 2 0 -2
χ' (× 10⁻⁴ emu/g)
Temperature (K)
Temperature (K)
32.6 32.8 33.0 33.2 33.4
(b)
expt. points
Dynamic fitted curve
Frequency (Hz)
700 600 500 400 300
(c)
Frequency (Hz)
Expt. points
V-F fitted curve
32.6 32.8 33.0 33.2 33.4
Temperature (K)
(d)
Experimental data
fit data
@25 K
m (t)
1.0010 1.0005 1.0000 0.9995 0.9990 0.9985 0.9980
0 2000 4000 6000 8000
t (sec)

**Figure. 7**

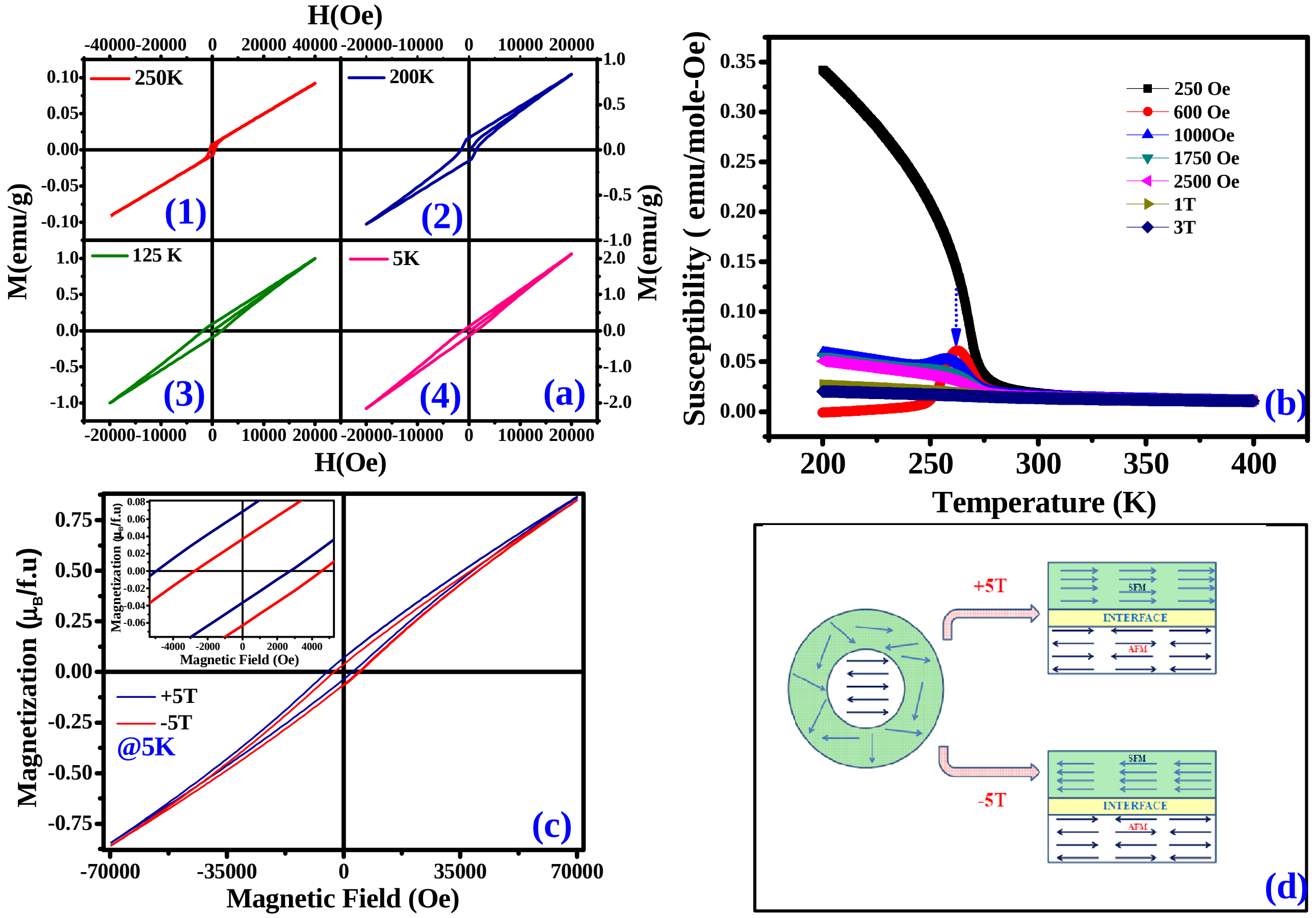

Figure. 8